\documentstyle[preprint,aps,epsf,floats]{revtex}

\tighten

\def\OMIT#1{{}}

\def\Dslash{D\hskip-0.65em /}
\def\Se{s}
\def\Tr{t}
\def\SD{x}
\def\SS{s}
\def\ST{t}
\def\SD{x}
\def\VS{\tilde s}
\def\VT{\tilde t}
\def\VD{\tilde x}
\def\an{\alpha^{(n)} }
\def\bn{\beta^{(n)} }
\def\gn{\gamma^{(n)} }
\def\sn{\sigma^{(n)} }
\def\cb{{\cal B}}
\def\cbb{{\overline{\cal B}}}
\def\cf{{\cal F}}

\begin{document}

\preprint{\vbox{
\hbox{NT@UW-01-026}
\hbox{UMPP\# 02-021}
}}

\title{Baryons 
in Partially Quenched Chiral Perturbation Theory}
\author{{\bf Jiunn-Wei Chen}$^a$ and {\bf Martin J. Savage}$^{b}$}
\address{$^a$ Department of Physics, University of Maryland, \\
College Park, MD 20742-4111.}
\address{$^b$ Department of Physics, University of Washington, \\
Seattle, WA 98195. }

\maketitle

\begin{abstract} 
We include the lowest-lying octet- and decuplet-baryons into partially
quenched chiral perturbation theory.
Perturbing about the chiral limit of the graded 
$SU(6|3)_L\otimes SU(6|3)_R$ flavor group of partially quenched QCD,
we compute the leading one-loop contributions to the octet-baryon masses,
magnetic moments and matrix elements of isovector twist-2 operators.
We work in the isospin limit and keep two of the three sea quarks degenerate.
The usefulness of the 
non-unique extension of the electric charge matrix and the isovector
twist-2 operators from QCD to partially quenched QCD  is discussed.
\end{abstract}

\bigskip
\vskip 8.0cm
\leftline{November 2001}

\vfill\eject

\section{Introduction}

Understanding and computing the properties and decays of hadrons 
remains the most significant challenge
presented by quantum chromodynamics (QCD), 
the theory of strong interactions.
While hadronic models of varying quality can describe 
a number of  these observables at some level of accuracy, 
a calculation of any observable (that is not a conserved charge)
directly from QCD is yet to be performed.
At some point in the future 
it will be possible to numerically evaluate these 
observables with lattice QCD.
While progress toward this ultimate goal is impressive, one is presently
restricted to lattice light quark masses, $m_q^{\rm latt}$ 
that are significantly larger than those of nature, $m_q$,
with typical pion masses being of order 
$m_\pi^{\rm latt}\sim 500~{\rm MeV}$.
Therefore, 
at present and in the foreseeable future, the $m_q$-dependence of observables
of interest need to be known in order to make a comparison between
lattice QCD results and experiment.

Chiral perturbation theory ($\chi$PT) can be used to
extrapolate unquenched calculations from $m_q^{\rm latt}$ to $m_q$
for small quark masses.
Further, if the masses are sufficiently small, the chiral expansion
will converge at low orders, making analytic calculation relatively
straightforward.
However, for the  $m_\pi^{\rm latt}$'s that are currently being
simulated, the convergence of the chiral expansion is somewhat uncertain,
to say the least, and high-order calculations (beyond those that currently
exist) are highly desirable.
Calculations of several of the observables have been performed 
in quenched QCD (QQCD)~\cite{S92,BG92,LS96}, 
in which the  contribution of the quark determinant to observables
is not evaluated,
reducing the time necessary for computation.
Unfortunately, such calculations cannot be connected to observables in QCD, 
and in many cases are found to be more divergent in the chiral limit
than those in QCD.
Recently, partially quenched QCD 
(PQQCD)~\cite{Pqqcda,Pqqcdb,Pqqcdc,Pqqcdd,Pqqcde,SS01}
has been formulated 
where the quarks that couple to external sources 
for the asymptotic hadron states, the valence-quarks,
are distinguished from those that contribute to the 
quark determinant, the sea-quarks.  
The main advantage that PQQCD enjoys is that
the masses of the valence quarks can be 
significantly smaller than those of the sea-quarks.
In order to perform the extrapolation in the valence and sea quark masses,
one employs partially quenched chiral perturbation theory 
(PQ$\chi$PT)~\cite{Pqqcda,Pqqcdb,Pqqcdc,Pqqcdd,Pqqcde,SS01}
to systematically expand about the chiral limit,
assuming that the quark masses, $m_Q$~\footnote{
In this context, $m_Q$ is used to denote the mass of one of the valence-
or sea-quarks, while $m_q$ is reserved for the physical masses 
of the three light valence-quarks $u$, $d$ and $s$.},
are small enough for such an expansion to converge at relatively low-orders.
It is easy to convince oneself that the counterterms in the 
PQ$\chi$PT Lagrangian with three sea-quarks
are related to those in the $\chi$PT Lagrangian describing 
QCD.  Thus, fitting the counterterms in PQ$\chi$PT allows one to make
QCD predictions at the physical values of the quark masses.

Naively, it would appear that computing observables in the low-energy 
effective field theories (EFT) 
would not be of much use, after all, there are
counterterms that must be determined.
If numerical simulations will fix the counterterms then why not 
compute the complete amplitude?
For the present values of $m_q^{\rm latt}$,
numerically 
separating the terms that are non-analytic in $m_q$ from those that are 
analytic in $m_Q$ is difficult.
However, for the smaller values of $m_Q$ encountered in the 
extrapolation, it is the non-analytic
terms that formally dominate the $m_Q$-dependence.
EFT allows one to compute these non-analytic
terms, thereby removing this ambiguity, 
and allowing for 
a significantly more reliable extrapolation to small $m_Q$'s.

In this work we incorporate the lowest-lying octet and decuplet of baryons 
into PQ$\chi$PT~\footnote{
Ref~\cite{CR97} has included baryons in PQQCD in the 
large-$N_c$ limit in an effort to continuously move from
QQCD to QCD.  Our work has little overlap with Ref~\cite{CR97}.
}.
We then compute the ${\cal O}\left(m_Q^{3/2}\right)$ 
contributions to the octet-baryon masses, the  
${\cal O}\left(m_Q^{1/2}\right)$ contributions to the magnetic moments of the 
octet-baryons and compute the  ${\cal O}\left(m_Q\log m_Q\right)$ 
contributions
to the matrix elements of the isovector twist-2 operators that
give the moments of the isovector parton distributions.
The computations are performed with 
three valence-quarks, three ghost-quarks and three sea-quarks
in the isospin limit. That is to say that
two valence-quarks are degenerate with each other and with 
two ghost-quarks. In addition two sea-quarks are degenerate with each other,
but have masses different from the ghost- and valence-quarks.

\section{PQ$\chi$PT}

The lagrange density of PQQCD is
\begin{eqnarray}
{\cal L} & = & 
\sum_{a,b=u,d,s} \overline{q}_V^a
\left[\ i\Dslash -m_{q}\ \right]_a^b\ q_{V,b}
\ +\ 
\sum_{\tilde a,\tilde b=\tilde u, \tilde d,\tilde s}
\overline{\tilde q}^{\tilde a}
\left[\ i\Dslash -m_{\tilde q}\ 
\right]_{\tilde a}^{\tilde b}\tilde q_{\tilde b}
\ +\ 
\sum_{a,b=j,l,r} \overline{q}_{\rm sea}^a
\left[\ i\Dslash -m_{\rm sea}\ \right]_a^b\ q_{{\rm sea}, b}
\nonumber\\
& = & 
\sum_{k,n=u,d,s,\tilde u, \tilde d,\tilde s,j,l,r} 
\overline{Q}^k\ 
\left[\ i\Dslash -m_{Q}\ \right]_k^n\ Q_n
\ \ \ ,
\label{eq:PQQCD}
\end{eqnarray}
where the $q_V$ are the three light valence-quarks, $u$, $d$, and $s$,
the $\tilde q$ are three light bosonic (ghost) quarks  
$\tilde u$, $\tilde d$, and $\tilde s$,
and the $q_{{\rm sea}}$ are the three sea-quarks $j$, $l$ and $r$.
The left- and right-handed 
valence-, sea-, and ghost-quarks are combined into
column vectors
\begin{eqnarray}
Q_L & = & \left(u,d,s,j,l,r,\tilde u,\tilde d,\tilde s\right)^T_L
\ \ ,\ \ 
Q_R \ =\  \left(u,d,s,j,l,r,\tilde u,\tilde d,\tilde s\right)^T_R
\ \ \ ,
\label{eq:quarkvec}
\end{eqnarray}
where the graded equal-time commutation relation for two fields is
\begin{eqnarray}
Q_i^\alpha ({\bf x}) Q_k^{\beta \dagger} ({\bf y}) - 
(-)^{\eta_i\eta_k}Q_k^{\beta \dagger} ({\bf y})Q_i^\alpha ({\bf x})
& = & 
\delta^{\alpha\beta}\delta_{ik}\delta^3({\bf x}-{\bf y})
\ \ \ ,
\label{eq:comm}
\end{eqnarray}
where $\alpha,\beta$ are spin-indices and $i,k$ are flavor indices.
The objects $\eta_k$ correspond to the parity of the component of $Q_k$,
with $\eta_k=+1$ for $k=1,2,3,4,5,6$ and $\eta_k=0$ for $k=7,8,9$, and 
the graded equal-time 
commutation relations for two $Q$'s or two $Q^\dagger$'s
are analogous.
The $Q_{L,R}$ in eq.~(\ref{eq:quarkvec})
transform in the fundamental representation of $SU(6|3)_{L,R}$
respectively.
The fermionic components of the left-handed field
$Q_L$ transform as a $({\bf 6},{\bf 1})$ of
$SU(6)_{qL}\otimes SU(3)_{\tilde q L}$ while the
bosonic components transform
as $({\bf 1},{\bf 3})$, and the 
right-handed field $Q_R$ transforms analogously.

In the absence of quark masses, $m_Q=0$,
the lagrange density in eq.~(\ref{eq:PQQCD})
has a
graded symmetry $U(6|3)_L\otimes U(6|3)_R$, where the left- and 
right-handed quark fields transform as
$Q_L\rightarrow U_L Q_L$ and $Q_R\rightarrow U_R Q_R$ respectively.
The strong anomaly reduces the symmetry of the theory to
$SU(6|3)_L\otimes SU(6|3)_R\otimes U(1)_V$~\cite{SS01}~\footnote{
This differs from the group structure of QQCD~\cite{SS01}.},
and it is assumed that this symmetry is spontaneously broken 
$SU(6|3)_L\otimes SU(6|3)_R\otimes U(1)_V\rightarrow 
SU(6|3)_V\otimes U(1)_V$ so that an identification with QCD can be made.

The mass-matrix, $m_Q$, has entries
$m_Q = {\rm diag}(m_u,m_d,m_s,m_j,m_l,m_r,m_u,m_d,m_s)$,
i.e.  
$m_{\tilde u}=m_u$, $m_{\tilde d}=m_d$ and 
$m_{\tilde s}=m_s$, so that the contribution 
to the determinant in the path integral 
from integrating over 
the $q$'s and the $\tilde q$'s exactly cancel, leaving the 
contribution from the $q_{\rm sea}$'s alone.
In the isospin limit,
\begin{eqnarray}
m_Q & = & {\rm diag}(\ \overline{m} , \overline{m} ,
m_s , m_j , m_j , m_r , \overline{m} , \overline{m} , m_s\ )
\label{eq:massmat}
\ \ \ .
\end{eqnarray}

\subsection{The Pseudo-Goldstone Bosons}

The strong interaction dynamics of the 
pseudo-Goldstone bosons are described  at leading order
in PQ$\chi$PT by a Lagrange density of the 
form~\cite{Pqqcda,Pqqcdb,Pqqcdc,Pqqcdd,Pqqcde},
\begin{eqnarray}
{\cal L } & = & 
{f^2\over 8} {\rm
  str}\left[\ \partial^\mu\Sigma^\dagger\partial_\mu\Sigma\ \right]
 \ +\ \lambda\ {\rm str}\left[\ m_Q\Sigma^\dagger + m_Q\Sigma\ \right]
\ +\ 
\alpha_\Phi\partial^\mu\Phi_0\partial_\mu\Phi_0\ -\ m_0^2\Phi_0^2
\ \ \ \ ,
\label{eq:lagpi}
\end{eqnarray}
where $\alpha_\Phi$ and $m_0$ are quantities that do not vanish in the 
chiral limit.
The operation ``${\rm str}$'' in eq.~(\ref{eq:lagpi}) is defined to
be the supertrace.
The meson field is incorporated in $\Sigma$ via
\begin{eqnarray}
\Sigma & = & \exp\left({2\ i\ \Phi\over f}\right)
\ =\ \xi^2
\ \ \ ,\ \ \ 
\Phi \ =\  \left(\matrix{ M &\chi^\dagger \cr \chi &\tilde{M} }\right)
\ \ \ ,
\label{eq:phidef}
\end{eqnarray}
where $M$ and $\tilde M$ are matrices containing bosonic mesons while
$\chi$ and $\chi^\dagger$ are matrices containing fermionic mesons,
with
\begin{eqnarray}
M & = & 
\left(\matrix{
\eta_u & \pi^+ & K^+ & J^0 & L^+ & R^+\cr
\pi^- & \eta_d & K^0 & J^- & L^0 & R^0\cr
K^- & \overline{K}^0 & \eta_s & J_s^- & L_s^0 & R_s^0\cr
\overline{J}^0 & J^+ & J_s^+ & \eta_j & Y_{jl}^+ & Y_{jr}^+\cr
L^- & \overline{L}^0 & \overline{L_s}^0 & Y_{jl}^- & \eta_l & Y_{lr}^0\cr
R^- & \overline{R}^0 & \overline{R_s}^0 & Y_{jr}^- & \overline{Y}_{lr}^0 & \eta_r }
\right)
\ \ \ ,\ \ \ 
\tilde M \ =\  \left(\matrix{\tilde\eta_u & \tilde\pi^+ & \tilde K^+\cr
\tilde\pi^- & \tilde\eta_d & \tilde K^0\cr
\tilde K^- & \tilde{\overline{K}^0} & \tilde\eta_s
}\right)
\nonumber\\
\chi & = & 
\left(\matrix{\chi_{\eta_u} & \chi_{\pi^+} & \chi_{K^+} & 
\chi_{J^0} & \chi_{L^+} & \chi_{R^+}\cr
\chi_{\pi^-} & \chi_{\eta_d} & \chi_{K^0} &
\chi_{J^-} & \chi_{L^0} & \chi_{R^0}\cr
\chi_{K^-} & \chi_{\overline{K}^0} & \chi_{\eta_s} & 
\chi_{J_s^-} & \chi_{L_s^0} & \chi_{R_s^0} }
\right)
\ \ \ \ .
\label{eq:mesdef}
\end{eqnarray}
where the upper $3\times 3$ block of $M$ is the usual octet of
pseudo-scalar mesons while the remaining entries correspond to mesons
formed with the sea-quarks.
The convention we use corresponds to $f~\sim~132~{\rm MeV}$.

The singlet field is defined to be 
$\Phi_0 ={\rm str}\left(\ \Phi\ \right)/\sqrt{2}$,
and its mass $m_0$ can be taken to
be of order the scale of chiral symmetry breaking, 
$m_0\rightarrow\Lambda_\chi$~\cite{SS01}.
In taking this limit, one finds that the $\eta$
two-point functions deviate from the simple, single pole form.
The $\eta_a\eta_b$ propagator for $2+1$ sea-quarks and $a, b = u, d, s$
at leading order,
is found to be
\begin{eqnarray}
{\cal G}_{\eta_a\eta_b} & = & 
{ i \delta^{ab}\over q^2- m_{\eta_a}^2 + i \epsilon}
\ -\ {i\over 3}
{(q^2-m_{jj}^2)(q^2-m_{rr}^2)\over (q^2- m_{\eta_a}^2 + i \epsilon)
(q^2- m_{\eta_b}^2 + i \epsilon)
(q^2- {1\over 3}(m_{jj}^2+2 m_{rr}^2)+ i \epsilon)}
\ \ ,
\end{eqnarray}
where $m_{xy}$ is the mass of the meson composed of (anti)-quarks of flavor
$x$ and $y$.
This can be compactly written as
\begin{eqnarray}
{\cal G}_{\eta_a\eta_b} & = &  \delta^{ab}  P_a\ +\ 
{\cal H}_{ab}( P_a , P_b , P_X)
\ \ ,
\label{eq:HPs}
\end{eqnarray}
where 
\begin{eqnarray}
P_a & = & { i \over q^2- m_{\eta_a}^2 + i \epsilon}
\ \ \ ,\ \ \ 
P_b \ = \  { i \over q^2- m_{\eta_b}^2 + i \epsilon}
\ \ \ ,\ \ \ 
P_X \ = \  { i \over q^2- m_X^2 + i \epsilon}
\nonumber\\
{\cal H}_{ab}( A, B, C) & = & 
-{1\over 3}\left[\ 
{(m_{jj}^2-m_{\eta_a}^2)(m_{rr}^2-m_{\eta_a}^2)\over 
(m_{\eta_a}^2-m_{\eta_b}^2)(m_{\eta_a}^2-m_X^2)}\  A
-
{(m_{jj}^2-m_{\eta_b}^2)(m_{rr}^2-m_{\eta_b}^2)\over 
(m_{\eta_a}^2-m_{\eta_b}^2)(m_{\eta_b}^2-m_X^2)}\  B
\right.\nonumber\\ & & \left.\qquad
\ +\ 
{(m_X^2-m_{jj}^2)(m_X^2-m_{rr}^2)\over 
(m_X^2-m_{\eta_a}^2)(m_X^2-m_{\eta_b}^2)}\  C
\ \right]
\ \ \ ,
\label{eq:HPsdef}
\end{eqnarray}
where the mass, $m_X$, is given by 
$m_X^2 = {1\over 3}\left( m_{jj}^2+2 m_{rr}^2 \right)$.

\subsection{The Baryons}

The method for including the lowest-lying baryons, 
the octet of spin-${1\over 2}$ baryons 
and the decuplet of spin-${3\over 2}$ baryon resonances,
into PQ$\chi$PT is similar to the method used
to include them into QQCD, as detailed in Ref.~\cite{LS96}.
An interpolating field that has non-zero overlap with the 
baryon octet (when the $ijk$ indices are restricted to $1,2,3$) 
is~\cite{LS96}
\begin{eqnarray}
{\cal B}^\gamma_{ijk} & \sim &
\left[\ Q_i^{\alpha,a} Q_j^{\beta,b} Q_k^{\gamma,c}
\ -\  Q_i^{\alpha,a} Q_j^{\gamma,c} Q_k^{\beta,b}\ \right]
\epsilon_{abc} \left(C\gamma_5\right)_{\alpha\beta}
\ \ \ ,
\label{eq:octinter}
\end{eqnarray}
where $C$ is the charge conjugation operator,
$a,b,c$ are color indices and $\alpha,\beta,\gamma$ are Dirac indices.
Dropping the Dirac index, one finds that
under the interchange of flavor indices~\cite{LS96},
\begin{eqnarray}
{\cal B}_{ijk} & = & (-)^{1+\eta_j \eta_k}\  {\cal B}_{ikj}
\ \ ,\ \ 
{\cal B}_{ijk} \ +\  (-)^{1+\eta_i \eta_j}\ {\cal B}_{jik}
\ +\ (-)^{1 + \eta_i\eta_j + \eta_j\eta_k + \eta_k\eta_i}\ 
{\cal B}_{kji}\ =\ 0
\ \ \ .
\label{eq:bianchi}
\end{eqnarray}
In analogy with QCD, 
we consider the transformation of $B_{ijk}$ under $SU(6|3)_V$
transformations, and using the graded relation
\begin{eqnarray}
Q_i\ U^j_{\ k} & = & (-)^{\eta_i (\eta_j+\eta_k)} \ U^j_{\ k}\  Q_i
\ \ ,
\label{eq:gradtrans}
\end{eqnarray}
in eq.~(\ref{eq:octinter}),
it is straightforward to show that~\cite{LS96}
\begin{eqnarray}
{\cal B}_{ijk} & \rightarrow & 
(-)^{\eta_l (\eta_j+\eta_m) +(\eta_l+\eta_m)(\eta_k+\eta_n)} 
\ U_i^{\ l}\  U_j^{\ m}\  U_k^{\ n}\ 
{\cal B}_{lmn}
\ \ \ .
\label{eq:octtrans}
\end{eqnarray}
${\cal B}_{ijk}$ describes a {\bf 240} dimensional representation
of $SU(6|3)_V$.
It is convenient to decompose the irreducible representations of $SU(6|3)_V$ 
into  irreducible representations
$SU(3)_{\rm val}\otimes SU(3)_{\rm sea}\otimes SU(3)_{\tilde q}\otimes U(1)
$~\cite{BBI81,BB81,HM83}, 
and we will forget about the $U(1)$'s from now on.
The subscript denotes where the $SU(3)$ acts, either on the valence $q$'s,
on the sea $q$'s, or on the $\tilde q$'s.
In order to locate a particular baryon in the irreducible representation
we employ the terminology, ground floor, first floor, second floor 
and so on, as it is common in the description of super-algebra multiplets.
The ground floor contains all the baryons that do not contain a  
bosonic quark,
the first floor contains all baryons that contain only one bosonic quarks,
the second floor contains all baryons that contain two bosonic quarks,
and the third floor contain the baryons that are comprised entirely of
bosonic quarks.
As a way of distinguishing between baryons containing some number of 
valence and sea quarks, we introduce ``levels''.
Level A is comprised of baryons that do not contain sea quarks,
level B is comprised of baryons containing one sea quarks,
level C is comprised of baryons containing two sea quarks,
and level D  is comprised of baryons composed only of sea quarks.

The ground floor of level A of the {\bf 240}-dimensional 
representation contains baryons
that are comprised of three valence quarks, $q_V q_V q_V$, and is therefore an 
$({\bf 8},{\bf 1},{\bf 1})$ of  
$SU(3)_{\rm val}\otimes SU(3)_{\rm sea}\otimes SU(3)_{\tilde q}$.
The octet-baryons are embedded as~\cite{LS96}
\begin{eqnarray}
{\cal B}_{abc} & = & {1\over\sqrt{6}}
\left( \ \epsilon_{abd}\ B^d_c\ +\ 
\epsilon_{acd} B^d_b\ \right)
\ \ \ ,
\end{eqnarray}
where the indices are restricted to take the values $a,b,c=1,2,3$ only.
The octet-baryon matrix is
\begin{eqnarray}
B & = & \left(\matrix{{1\over\sqrt{6}}\Lambda + {1\over\sqrt{2}}\Sigma^0
& \Sigma^+ & p\cr
\Sigma^- & {1\over\sqrt{6}}\Lambda - {1\over\sqrt{2}}\Sigma^0 & n\cr
\Xi^- & \Xi^0 & -{2\over\sqrt{6}}\Lambda}\right)
\ \ \ .
\label{eq:baryons}
\end{eqnarray}
The first floor of level A of the {\bf 240}-dimensional representation 
contains baryons that are composed of two valence quarks and 
one ghost-quark, $\tilde q q_V q_V$, and therefore transforms as 
$({\bf 6}, {\bf 1}, {\bf 3})\oplus (\overline{\bf 3}, {\bf 1}, {\bf 3})$
of $SU(3)_{\rm val}\otimes SU(3)_{\rm sea}\otimes SU(3)_{\tilde q}$ .
The tensor representation $_{\tilde a} \tilde\Se_{ab}$ 
of the $({\bf 6}, {\bf 1}, {\bf 3})$ multiplet, 
where ${\tilde a}=1,2,3$ runs over the 
$\tilde q$ indices and $a,b=1,2,3$ run over the $q_V$ indices,
has baryon assignment
\begin{eqnarray}
_{\tilde a} \tilde\Se_{11} & = & 
\tilde\Sigma_{\tilde a}^{+1}
\ \ ,\ \ 
_{\tilde a} \tilde\Se_{12}\ =\ _{\tilde a} \tilde\Se_{21}\ =\ 
{1\over\sqrt{2}} \tilde\Sigma_{\tilde a}^{0}
\ \ ,\ \ 
_{\tilde a} \tilde\Se_{22}\ =\ 
\tilde\Sigma_{\tilde a}^{-1}
\nonumber\\
_{\tilde a} \tilde\Se_{13}& = & _{\tilde a} \tilde\Se_{31}
\ = \
{1\over\sqrt{2}}
\ ^{(6)}\tilde\Xi_{\tilde a}^{+{1\over 2}}
\ \ ,\ \ 
_{\tilde a} \tilde\Se_{23}\ =\ _{\tilde a} \tilde\Se_{32}
\ =\ 
{1\over\sqrt{2}}
\ ^{(6)}\tilde\Xi_{\tilde a}^{-{1\over 2}}
\ \ ,\ \ 
_{\tilde a} \tilde\Se_{33} \ =\ 
\tilde\Omega_{\tilde a}^0
\ \ \ ,
\label{eq:sixdef}
\end{eqnarray}
The right superscript denotes the third component of $q_V$-isospin,
while the left subscript denotes the $\tilde q$ flavor.
The tensor representation $_{\tilde a} \tilde\Tr^a$ 
of the $(\overline{\bf 3},{\bf 1}, {\bf 3})$ 
multiplet, where ${\tilde a}=1,2,3$ runs over the 
$\tilde q$ indices and $a=1,2,3$ run over the $q_V$ indices,
has baryon assignment
\begin{eqnarray}
_{\tilde a}\tilde \Tr^1 & = & ^{(\overline{3})}
\tilde\Xi_{\tilde a}^{-{1\over 2}}
\ \ ,\ \ 
_{\tilde a} \tilde\Tr^2 \ =\  ^{(\overline{3})}
\tilde\Xi_{\tilde a}^{+{1\over 2}}
\ \ ,\ \ 
_{\tilde a} \tilde\Tr^3 \ =\  \tilde\Lambda_{\tilde a}^0
\ \ \ .
\label{eq:tripdef}
\end{eqnarray}
The ground floor of level B of the {\bf 240}-dimensional representation 
contains baryons that are composed of two valence quarks and 
one sea quark, $q_V q_V q_{\rm sea}$, and therefore transforms as 
$({\bf 6}, {\bf 3}, {\bf 1})\oplus (\overline{\bf 3}, {\bf 3}, {\bf 1})$
of $SU(3)_{\rm val}\otimes SU(3)_{\rm sea}\otimes SU(3)_{\tilde q}$ .
The tensor representation $_a \Se_{bc}$ 
of the $({\bf 6}, {\bf 3}, {\bf 1})$ multiplet, 
where ${a}=1,2,3$ runs over the 
$q_{\rm sea}$ indices and $b,c=1,2,3$ run over the $q_V$ indices,
has baryon assignment
\begin{eqnarray}
_{a} \Se_{11} & = & 
\Sigma_{a}^{+1}
\ \ ,\ \ 
_{a} \Se_{12}\ =\ _{a}\Se_{21}\ =\ 
{1\over\sqrt{2}} \Sigma_{a}^{0}
\ \ ,\ \ 
_{a}\Se_{22}\ =\ \Sigma_{a}^{-1}
\nonumber\\
_{a}\Se_{13}& = & _{a} \Se_{31}
\ = \
{1\over\sqrt{2}}
\ ^{(6)}\Xi_{a}^{+{1\over 2}}
\ \ ,\ \ 
_{a}\Se_{23}\ =\ _{a} \Se_{32}
\ =\ 
{1\over\sqrt{2}}
\ ^{(6)}\Xi_{a}^{-{1\over 2}}
\ \ ,\ \ 
_{a} \Se_{33} \ =\ \Omega_{a}^0
\ \ \ .
\label{eq:sixdefsea}
\end{eqnarray}
The tensor representation $_{a} \Tr^b$ 
of the $(\overline{\bf 3}, {\bf 3}, {\bf 1})$ 
multiplet, where ${a}=1,2,3$ runs over the 
$q_{\rm sea}$ indices and $b=1,2,3$ run over the $q$ indices,
has baryon assignment
\begin{eqnarray}
_{a} \Tr^1 & = & ^{(\overline{3})}\Xi_{a}^{-{1\over 2}}
\ \ ,\ \ 
_{a} \Tr^2 \ =\  ^{(\overline{3})}\Xi_{a}^{+{1\over 2}}
\ \ ,\ \ 
_{a} \Tr^3 \ =\  \Lambda_{a}^0
\ \ \ .
\label{eq:tripdefsea}
\end{eqnarray}

The $_{\tilde a} \tilde\Se_{ab}$, $_{\tilde a}  \tilde\Tr^a$ 
$_{a}\Se_{ab}$, and $_{a}\Tr^a$ 
are uniquely embedded into
${\cal B}_{ijk}$ (up to field redefinition's), 
constrained by the relations in eq.~(\ref{eq:bianchi}):
\begin{eqnarray}
{\cal B}_{ijk} & = & 
-\sqrt{2\over 3\ }\ _{i-3}\SS_{jk}
\ \ \ \ \ {\rm for}\  \ \ \ i=4,5,6\ \ {\rm and}\ \ j,k=1,2,3
\nonumber\\
{\cal B}_{ijk} & = & 
{1\over 2}\ \  _{j-3}\ST^\sigma \varepsilon_{\sigma i k }
\ +\ {1\over\sqrt{6}}\ \ _{j-3}\SS_{ik}
\ \ \ \ \ {\rm for}\  \ \ \ j=4,5,6\ \ {\rm and}\ \ i,k,\sigma =1,2,3
\nonumber\\
{\cal B}_{ijk} & = & 
{1\over 2}\ \   _{k-3}\ST^\sigma \varepsilon_{\sigma i j }
\ +\ {1\over\sqrt{6}}\ \  _{k-3}\SS_{ij}
\ \ \ \ \ {\rm for}\  \ \ \ k=4,5,6\ \ {\rm and}\ \ i,j,\sigma =1,2,3
\nonumber\\
{\cal B}_{ijk} & = & 
\sqrt{2\over 3\ }\ _{i-6}\VS_{jk}
\ \ \ \ \ {\rm for}\  \ \ \ i=7,8,9\ \ {\rm and}\ \ j,k=1,2,3
\nonumber\\
{\cal B}_{ijk} & = & 
{1\over 2}\ \  _{j-6}\VT^\sigma \varepsilon_{\sigma i k }
\ +\ {1\over\sqrt{6}}\ \ _{j-6}\VS_{ik}
\ \ \ \ \ {\rm for}\  \ \ \ j=7,8,9\ \ {\rm and}\ \ i,k,\sigma =1,2,3
\nonumber\\
{\cal B}_{ijk} & = & 
-{1\over 2}\ \   _{k-6}\VT^\sigma \varepsilon_{\sigma i j }
\ -\ {1\over\sqrt{6}}\ \  _{k-6}\VS_{ij}
\ \ \ \ \ {\rm for}\  \ \ \ k=7,8,9\ \ {\rm and}\ \ i,j,\sigma =1,2,3
\ \ \ .
\label{eq:embedoctet}
\end{eqnarray}
As we are only interested in one-loop contributions to observables with 
$q_V q_V q_V$-baryons in the asymptotic states, we do not 
explicitly construct the remaining floors and levels of the {\bf 240}.


As the mass splitting between the 
decuplet- and octet-baryons
(in QCD) is much less than the scale of chiral
symmetry breaking ($\Lambda_\chi\sim 1~{\rm GeV}$)
the decuplet must be included as a dynamical field in order to have a
theory where the natural scale of higher order interactions is set by
$\Lambda_\chi$.
We assume that the decuplet-octet mass splitting, $\Delta$, 
remains small compared to the 
scale of chiral symmetry breaking in PQQCD.
An interpolating field that contains the spin-${3\over 2}$
decuplet as the ground floor of level A is~\cite{LS96}
\begin{eqnarray}
{\cal T}^{\alpha ,\mu}_{ijk} & \sim &
\left[
Q^{\alpha,a}_i Q^{\beta,b}_j Q^{\gamma,c}_k +
Q^{\beta,b}_i Q^{\gamma,c}_j Q^{\alpha,a}_k  +
Q^{\gamma,c}_i Q^{\alpha,a}_j Q^{\beta,b}_k 
\right]
\varepsilon_{abc} (C\gamma^\mu)_{\beta\gamma}
\ \ \ ,
\label{eq:tdef}
\end{eqnarray}
where the indices $i,j,k$ run from $1$ to $9$.
Neglecting Dirac indices, one finds that
under the interchange of flavor indices~\cite{LS96}
\begin{eqnarray}
{\cal T}_{ijk} & = & 
(-)^{1+\eta_i\eta_j} {\cal T}_{jik}\ =\ 
(-)^{1+\eta_j\eta_k} {\cal T}_{ikj}
\ \ \ .
\label{eq:ttrans}
\end{eqnarray}
${\cal T}_{ijk}$ describes a ${\bf 138}$ dimensional representation
of $SU(6|3)_V$, which has the ground floor of level A transforming as 
$({\bf 10}, {\bf 1}, {\bf 1})$ under 
$SU(3)_{\rm val} \otimes SU(3)_{\rm sea} \otimes SU(3)_{\tilde q}$ with
\begin{eqnarray}
{\cal T}_{abc} & = & T_{abc}
\ \ \ ,
\label{eq:Tbarys}
\end{eqnarray}
where the indices are restricted to take the values $a,b,c=1,2,3$,
and where $T_{abc}$ is the totally symmetric tensor containing
the decuplet of baryon resonances,
\begin{eqnarray}
T_{111} & = & \Delta^{++}
\ \ ,\ \ 
T_{112} \ =\  {1\over\sqrt{3}}\Delta^+
\ \ ,\ \ 
T_{122} \ =\  {1\over\sqrt{3}}\Delta^0
\ \ ,\ \ 
T_{222} \ =\   \Delta^{-}
\nonumber\\
T_{113} & = & {1\over\sqrt{3}}\Sigma^{*,+}
\ \ ,\ \ 
T_{123} \ =\  {1\over\sqrt{6}}\Sigma^{*,0}
\ \ ,\ \ 
T_{223} \ =\  {1\over\sqrt{3}}\Sigma^{*,-}
\nonumber\\
T_{133} & = & {1\over\sqrt{3}}\Xi^{*,0}
\ \ ,\ \ 
T_{233} \ =\  {1\over\sqrt{3}}\Xi^{*,-}
\ \ ,\ \ 
T_{333} \ =\  \Omega^{-}
\ \ \ .
\label{eq:decuplet}
\end{eqnarray}

The first floor of level A of the {\bf 138} 
transforms as a $({\bf 6},  {\bf 1}, {\bf 3})$ under
$SU(3)_{\rm val} \otimes SU(3)_{\rm sea} \otimes SU(3)_{\tilde q}$
which has a tensor representation,
$_{\tilde a} \tilde\SD_{ij}$, with baryon assignment
\begin{eqnarray}
_{\tilde a} \tilde\SD_{11} & = & 
\tilde\Sigma_{\tilde a}^{*,+1}
\ \ ,\ \ 
_{\tilde a} \tilde\SD_{12}\ =\ _{\tilde a}\tilde \SD_{21}\ =\ 
{1\over\sqrt{2}} \tilde\Sigma_{\tilde a}^{*,0}
\ \ ,\ \ 
_{\tilde a} \tilde\SD_{22}\ =\ 
\tilde\Sigma_{\tilde a}^{*,-1}
\nonumber\\
_{\tilde a} \tilde\SD_{13} & = & _{\tilde a} \tilde\SD_{31}
\ =\ 
{1\over\sqrt{2}}
\ \tilde\Xi_{\tilde a}^{*,+{1\over 2}}
\ \ ,\ \ 
_{\tilde a} \tilde\SD_{23}\ =\ _{\tilde a} \tilde\SD_{32}
\ =\ 
{1\over\sqrt{2}}
\ \tilde \Xi_{\tilde a}^{*,-{1\over 2}}
\ \ ,\ \ 
_{\tilde a}  \tilde\SD_{33} \ =\  
 \tilde\Omega_{\tilde a}^{*,0}
\ \ \ .
\label{eq:sixTdef}
\end{eqnarray}
Similarly,
the ground floor of level B of the {\bf 138} 
transforms as a $({\bf 6}, {\bf 3}, {\bf 1})$ under
$SU(3)_{\rm val} \otimes SU(3)_{\rm sea} \otimes SU(3)_{\tilde q}$
which has a tensor representation,
$_{a} \SD_{ij}$, with baryon assignment
\begin{eqnarray}
_{a} \SD_{11} & = & 
\Sigma_{a}^{*,+1}
\ \ ,\ \ 
_{a} \SD_{12}\ =\ _{a} \SD_{21}\ =\ 
{1\over\sqrt{2}} \Sigma_{a}^{*,0}
\ \ ,\ \ 
_{a} \SD_{22}\ =\ 
\Sigma_{a}^{*,-1}
\nonumber\\
_{a} \SD_{13} & = & _{a} \SD_{31}
\ =\ 
{1\over\sqrt{2}}
\ \Xi_{a}^{*,+{1\over 2}}
\ \ ,\ \ 
_{a} \SD_{23}\ =\ _{a} \SD_{32}
\ =\ 
{1\over\sqrt{2}}
\ \Xi_{a}^{*,-{1\over 2}}
\ \ ,\ \ 
_{ a} \SD_{33} \ =\  
\Omega_{a}^{*,0}
\ \ \ .
\label{eq:sixTdefsea}
\end{eqnarray}
The embedding of $_{\tilde a} \VD_{ij}$ and $_a \SD_{ij}$
into ${\cal T}_{ijk}$ is 
unique (up to field redefinition's),
constrained by the symmetry properties in eq.~(\ref{eq:ttrans}):
\begin{eqnarray}
{\cal T}_{ijk} & = & 
+ {1\over\sqrt{3}}\ _{i-3}\SD_{jk}
\ \ \ \ \ {\rm for}\  \ \ \ i=4,5,6\ \ {\rm and}\ \ j,k=1,2,3
\nonumber\\
{\cal T}_{ijk} & = & 
{1\over\sqrt{3}}\ \ _{j-3}\SD_{ik}
\ \ \ \ \ {\rm for}\  \ \ \ j=4,5,6\ \ {\rm and}\ \ i,k =1,2,3
\nonumber\\
{\cal T}_{ijk} & = & 
+ {1\over\sqrt{3}}\ \  _{k-3}\SD_{ij}
\ \ \ \ \ {\rm for}\  \ \ \ k=4,5,6\ \ {\rm and}\ \ i,j =1,2,3
\nonumber\\
{\cal T}_{ijk} & = & 
+ {1\over\sqrt{3}}\ _{i-6}\tilde \SD_{jk}
\ \ \ \ \ {\rm for}\  \ \ \ i=7,8,9\ \ {\rm and}\ \ j,k=1,2,3
\nonumber\\
{\cal T}_{ijk} & = & 
-{1\over\sqrt{3}}\ \ _{j-6}\tilde\SD_{ik}
\ \ \ \ \ {\rm for}\  \ \ \ j=7,8,9\ \ {\rm and}\ \ i,k =1,2,3
\nonumber\\
{\cal T}_{ijk} & = & 
+ {1\over\sqrt{3}}\ \  _{k-6}\tilde\SD_{ij}
\ \ \ \ \ {\rm for}\  \ \ \ k=7,8,9\ \ {\rm and}\ \ i,j =1,2,3
\ \ \ .
\label{eq:firstfloorT}
\end{eqnarray}
We do not explicitly
construct the second and third floor baryons of the {\bf 138}
as we will only compute one-loop diagrams with octet-baryons 
in the asymptotic states.

\subsection{Lagrange Density for the Baryons}

The free Lagrange density for the ${\cal B}_{ijk}$ and 
${\cal T}_{ijk}$ fields is~\cite{LS96}, at leading order in the heavy baryon 
expansion~\cite{JMheavy,JMaxial,Jmass,chiralN,chiralUlf},
\begin{eqnarray}
{\cal L} & = & 
i\left(\overline{\cal B} v\cdot {\cal D} {\cal B}\right)
\ +\ 2\alpha_M \left(\overline{\cal B}{\cal B}{\cal M}_+\right)
\ +\ 2\beta_M \left(\overline{\cal B}{\cal M}_+{\cal B}\right)
\ +\ 2\sigma_M \left(\overline{\cal B}{\cal B}\right)\ 
{\rm str}\left({\cal M}_+\right)
\nonumber\\
& - & 
i \left(\overline{\cal T}^\mu v\cdot {\cal D} {\cal T}_\mu\right)
\ +\ 
\Delta\ \left(\overline{\cal T}^\mu {\cal T}_\mu\right)
\ +\ 2\gamma_M \left(\overline{\cal T}^\mu{\cal M}_+{\cal T}_\mu\right)
\ -\ 2 \overline{\sigma}_M  \left(\overline{\cal T}^\mu {\cal T}_\mu\right)\
{\rm str}\left({\cal M}_+\right)
\ \ ,
\label{eq:free}
\end{eqnarray}
where 
${\cal M}_+={1\over 2}\left(\xi^\dagger m_Q\xi^\dagger + \xi m_Q\xi\right)$,
and $\xi=\sqrt{\Sigma}$.
The brackets, $\left(\ \right)$ denote contraction of lorentz and flavor
indices as defined in Ref.~~\cite{LS96}.
For a matrix $\Gamma^\alpha_\beta$ acting in spin-space, 
and a matrix $Y_{ij}$ that acts in flavor-space, 
the required contractions are~\cite{LS96}
\begin{eqnarray}
\left(\overline{\cal B}\  \Gamma \ {\cal B}\right)
& = & 
\overline{\cal B}^{\alpha,kji}\ \Gamma_\alpha^\beta\  {\cal B}_{ijk,\beta}
\ \ ,\ \ 
\left(\overline{\cal T}^\mu\  \Gamma \ {\cal T}_\mu\right)
\ =\ 
\overline{\cal T}^{\mu\alpha,kji}\ \Gamma_\alpha^\beta\  
{\cal T}_{ijk,\beta\mu}
\nonumber\\
\left(\overline{\cal B}\  \Gamma \ Y\ {\cal B}\right)
& = & 
\overline{\cal B}^{\alpha,kji}\ \Gamma_\alpha^\beta\  
Y_i^{\ l}\ 
{\cal B}_{ljk,\beta}
\ \ ,\ \ 
\left(\overline{\cal T}^\mu\  \Gamma \ Y\ {\cal T}_\mu\right)
\ =\ 
\overline{\cal T}^{\mu\alpha,kji}\ \Gamma_\alpha^\beta\  
Y_i^{\ l}\ 
{\cal T}_{ljk,\beta\mu}
\nonumber\\
\left(\overline{\cal B}\  \Gamma \ {\cal B}\ Y\right)
& = & 
(-)^{(\eta_i+\eta_j)(\eta_k+\eta_n)}
\overline{\cal B}^{\alpha,kji}\ \Gamma_\alpha^\beta\  
Y_k^{\ n}\ 
{\cal B}_{ijn,\beta}
\nonumber\\
\left(\overline{\cal B}\  \Gamma \ Y^\mu {\cal T}_\mu\right)
& = & 
\overline{\cal B}^{\alpha,kji}\ \Gamma_\alpha^\beta\  
\left(Y^\mu\right)_i^l
{\cal T}_{ljk,\beta\mu}
\ \ \ ,
\label{eq:Contractions}
\end{eqnarray}
where $\overline{\cal B}$ and $\overline{\cal T}$
transform in the same way,
\begin{eqnarray}
\overline{B}^{kji}&\rightarrow &
(-)^{\eta_l (\eta_j+\eta_m) +(\eta_l+\eta_m)(\eta_k+\eta_n)} 
\overline{\cal B}^{nml}
 U_{n}^{\ k\dagger}\ U_{m}^{\ j\dagger}\ U_{l}^{\ i\dagger} \  
\ \ \ .
\end{eqnarray}

The Lagrange density describing the interactions of the baryons with the
pseudo-Goldstone bosons is~\cite{LS96}
\begin{eqnarray}
{\cal L} & = & 
2\alpha\ \left(\overline{\cal B} S^\mu {\cal B} A_\mu\right)
\ +\ 
2\beta\ \left(\overline{\cal B} S^\mu A_\mu {\cal B} \right)
\ +\  
2{\cal H} \left(\overline{\cal T}^\nu S^\mu A_\mu {\cal T}_\nu \right)
\nonumber\\
& &  
\ +\ 
\sqrt{3\over 2}{\cal C} 
\left[\ 
\left( \overline{\cal T}^\nu A_\nu {\cal B}\right)\ +\ 
\left({\cal B} A_\nu {\cal T}^\nu\right)\ \right]
\ ,
\label{eq:ints}
\end{eqnarray}
where $S^\mu$ is the covariant spin-vector~\cite{JMheavy,JMaxial,Jmass}.
Restricting oneself to the $q_V q_V q_V$ sector, it is straightforward to show
that 
\begin{eqnarray}
\alpha & =& {2\over 3}D+2F
\ \ ,\ \ 
\beta \ =\ -{5\over 3}D + F 
\ \ ,
\label{eq:couplings}
\end{eqnarray}
where $D$ and $F$ are constants that multiply the $SU(3)_{\rm val}$ 
invariants that are commonly used in QCD.
It should be stressed that the $F$ and $D$ discussed here in PQQCD 
are the same as those of QCD,
and consequently in our calculations we will
replace $\alpha$ and $\beta$ with $F$ and $D$.
In the above discussion, vector and axial-vector meson
fields have been introduced in analogy with QCD.
The covariant derivative acting on either the ${\cal B}$ or ${\cal T}$ fields
has the form
\begin{eqnarray}
\left({\cal D}^\mu{\cal B}\right)_{ijk} & = & 
\partial^\mu {\cal B}_{ijk}
+
\left(V^\mu\right)^l_i {\cal B}_{ljk}
+ 
(-)^{\eta_i (\eta_j+\eta_m)} \left(V^\mu\right)^m_j {\cal B}_{imk}
+ (-)^{(\eta_i+\eta_j) (\eta_k+\eta_n)}
\left(V^\mu\right)^n_k {\cal B}_{ijn}
\label{eq:covariant}
\end{eqnarray}
where the vector and axial-vector meson fields are
\begin{eqnarray}
V^\mu & = & {1\over 2}\left(\ \xi\partial^\mu\xi^\dagger
\ + \ 
\xi^\dagger\partial^\mu\xi \ \right)
\ \ ,\ \ 
A^\mu \ =\  {i\over 2}\left(\ \xi\partial^\mu\xi^\dagger
\ - \ 
\xi^\dagger\partial^\mu\xi \ \right)
\ \ \ .
\label{eq:mesonfields}
\end{eqnarray}

\section{Baryon Masses}

\begin{figure}[!ht]
\centerline{{\epsfxsize=4.0in \epsfbox{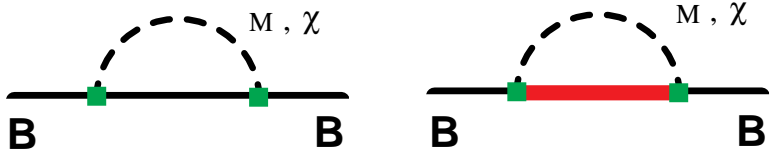}}} 
\vskip 0.15in
\noindent
\caption{\it 
One-loop graphs that give contributions of the form 
$\sim m_Q^{3/2}$
to the masses of the octet-baryons.
A solid, thick-solid and dashed line denote an 
{\bf 240}-baryon, {\bf 138}-baryon, and a meson, respectively.
The solid-squares denote an axial coupling given in eq.(\ref{eq:ints}).
}
\label{fig:masses}
\vskip .2in
\end{figure}

The masses of the octet-baryons provide nice example of how PQ$\chi$PT
can be implemented to determine coefficients in the QCD chiral lagrangian.
The mass of the $i$-th baryon has a chiral expansion
\begin{eqnarray}
M_i & = & M_0(\mu)\ -\ M_i^{(1)}(\mu)\ -\ M_i^{(3/2)}(\mu)\ +\ ...
\ \ \ ,
\label{eq:massexp}
\end{eqnarray}
where a term $M_i^{(n)}$ denotes a contribution of order $m_Q^n$. 
The baryon mass is dominated by a term in the PQ$\chi$PT Lagrange density, 
$M_0$, that is independent of $m_Q$.
Each of the contributions depend upon the scale chosen to 
renormalize the theory.
While at leading order (LO) and next-to-leading order (NLO)
in the chiral expansion, the objects $M_0$ and $ M_i^{(1)}$ are scale
independent, at one-loop level they are required to be scale dependent.
The leading dependence upon $m_Q$, occurring at order 
${\cal O}\left(m_Q\right)$, is due to the terms 
in eq.~(\ref{eq:free}) with coefficients $\alpha_M$, $\beta_M$ 
and $\sigma_M$, each of which need to be determined from lattice simulations.
The leading non-analytic dependence upon $m_Q$ arises from the one-loop
diagrams shown in Fig.~\ref{fig:masses}.
We find the contributions to the nucleon mass are
\begin{eqnarray}
M_N^{(1)} & = & \ 2\ \overline{m}\  (\alpha_M+\beta_M)\ +\ 
2\ \sigma_M\  (2m_j+m_r)
\nonumber\\
M_N^{(3/2)} & = & 
{1\over 8\pi f^2}\left[\ 
4D(F-{1\over 3}D)\  m_\pi^3 \ +\  
{1\over 3}(5D^2-6DF+9F^2)\ (2 m_{ju}^3+m_{ru}^3)
\right.\nonumber\\  & & \left.\qquad\qquad
+\ (D-3F)^2\ G_{\pi , \pi}
\ +\ {2 {\cal C}^2\over 3\pi}\left( F_{ju}\ +\ F_\pi \ +\ 
 {1\over 2} F_{ru}\right)
\ \right]
\ \ \ ,
\label{eq:Nmass}
\end{eqnarray}
where the function 
$G_{\pi ,\pi}= 
{\cal H}_{\pi\pi}(m_\pi^{3},m_\pi^{3},m_X^{3})$ 
is defined in eq.~(\ref{eq:HPsdef}).
The function $F_{c}=F( m_{c},\Delta,\mu)$ is
\begin{eqnarray}
F (m,\Delta,\mu) & = & 
\left(m^2-\Delta^2\right)\left(
\sqrt{\Delta^2-m^2} \log\left({\Delta -\sqrt{\Delta^2-m^2+i\epsilon}\over
\Delta +\sqrt{\Delta^2-m^2+i\epsilon}}\right)
-\Delta \log\left({m^2\over\mu^2}\right)\ \right)
\nonumber\\
& & \qquad -{1\over 2}\Delta m^2 \log\left({m^2\over\mu^2}\right)
\ \ \ .
\label{eq:massfun}
\end{eqnarray}
For the $\Sigma$'s we find
\begin{eqnarray}
M_\Sigma^{(1)} & = & 
\ {1\over 3}\ \overline{m}\  (5\alpha_M+2\beta_M)\ +\ 
\ {1\over 3}\ m_s\  (\alpha_M+4\beta_M)\ +\ 
\ 2 \sigma_M\  (2m_j+m_r)
\nonumber\\
M_\Sigma^{(3/2)} & = & 
{1\over 8\pi f^2}\left[\ 
-{2\over 3}(D^2-3F^2)\ m_\pi^3
\ -\ {2\over 3}(D^2-6DF+3F^2)\  m_K^3
\right.\nonumber\\ & & \left.
\qquad +\ (D-F)^2\  (2 m_{js}^3+m_{rs}^3) 
\ +\  {2\over 3}(D^2+3F^2)\ (2 m_{ju}^3+m_{ru}^3)
\right.\nonumber\\ & & \left.
\qquad +\ 4 F^2\  G_{\pi ,\pi} \ +\ 
 (D-F)^2 \  G_{\eta_s ,\eta_s}\  +\  4F(F-D)\ G_{\pi , \eta_s}
\right.\nonumber\\ & & \left.
\qquad +{2{\cal C}^2\over 3\pi}
\left(\ 
{5\over 6}F_K+{1\over 6}F_\pi+{1\over 3}( 2F_{js}+F_{rs})
+{1\over 6}(2F_{ju}+F_{ru})
\right.\right.\nonumber\\ & & \left.\left.\qquad\qquad
+{1\over 3}\left(\  E_{\pi ,\pi} + E_{\eta_s , \eta_s} - 2 E_{\pi , \eta_s}
\ \right)\ \right)
\ \right]
\ \ \ ,
\label{eq:Sigmass}
\end{eqnarray}
where 
$G_{\eta_s ,\eta_s} = 
{\cal H}_{\eta_s\eta_s}(m_{\eta_s}^{3},m_{\eta_s}^{3},m_X^{3})$
and 
$G_{\pi ,\eta_s} = {\cal H}_{\pi\eta_s}(m_\pi^{3},m_{\eta_s}^{3},m_X^{3})$.
The functions arising from loops involving decuplet intermediate states
are 
$E_{\pi ,\pi}={\cal H}_{\pi , \pi}(F_\pi, F_\pi, F_X)$, 
$E_{\eta_s ,\eta_s}={\cal H}_{\eta_s , \eta_s}(F_{\eta_s}, F_{\eta_s}, F_X)$,
and
$E_{\pi ,\eta_s}={\cal H}_{\pi , \eta_s}(F_{\pi}, F_{\eta_s}, F_X)$.
Contributions to the mass of the $\Lambda$ are found to be 
\begin{eqnarray}
M_\Lambda^{(1)} & = & 
\overline{m}\ (\alpha_M+2\beta_M)\ +\ m_s\ \alpha_M
\ +\ 
\ 2 \sigma_M\  (2m_j+m_r)
\nonumber\\
M_\Lambda^{(3/2)} & = & 
{1\over 8\pi f^2}\left[\ 
{2\over 9} (9F^2+6DF-5D^2)\ m_K^3 \ -\  
{2\over 9} (D^2-12 DF+9F^2)\  m_\pi^3
\right.\nonumber\\ & & \left.
+\ {1\over 9}(D+3F)^2\ (2 m_{js}^3+m_{rs}^3)
+{2\over 9} (7D^2-12 DF+9F^2)(2 m_{ju}^3+m_{ru}^3)
\right.\nonumber\\ & & \left.
+{4\over 9}(2D-3F)^2\  G_{\pi , \pi}
+{1\over 9}(D+3F)^2 \ G_{\eta_s , \eta_s}
-{4\over 9}(2D^2+3DF-9F^2)\ G_{\pi , \eta_s}
\right.\nonumber\\ & & \left.
+{{\cal C}^2\over 3\pi}
\left( F_K\ +\ F_\pi\ +\ F_{ru}\ +\ 2 F_{ju}\right)
\right]
\ \ \ .
\label{eq:lammass}
\end{eqnarray}
Finally, the contributions to the mass of the $\Xi$'s are found to be 
\begin{eqnarray}
M_\Xi^{(1)} & = & 
\ {1\over 3}\ \overline{m}\  (\alpha_M+4\beta_M)\ +\ 
\ {1\over 3}\ m_s\  (5\alpha_M+2\beta_M)\ +\ 
\ 2 \sigma_M\  (2m_j+m_r)
\nonumber\\
M_\Xi^{(3/2)} & = & 
{1\over 8\pi f^2}\left[\ 
-{2\over 3}(D^2-6DF+3F^2)\  m_K^3 
\ -\  {2\over 3} (D^2-3F^2) \ m_{\eta_s}^3
\right.\nonumber\\ & & \left.
\ +\ (D-F)^2 \ (2 m_{ju}^3+m_{ru}^3)
\ +\ {2\over 3}(D^2+3F^2)\ (2 m_{js}^3+m_{rs}^3)
\right.\nonumber\\ & & \left.
\ +\ (D-F)^2 \ G_{\pi , \pi}
\ +\ 4F(F-D) \ G_{\pi , \eta_s}
\ +\ 4F^2 \ G_{\eta_s , \eta_s}
\right.\nonumber\\ & & \left.
+{2{\cal C}^2\over 9\pi}\left(
{5\over 2} F_K+{1\over 2}F_{\eta_s}
+F_{js}+2 F_{ju}+{1\over 2} F_{rs}+F_{ru}
+E_{\pi , \pi} + E_{\eta_s , \eta_s} - 2 E_{\pi , \eta_s}
\right)
\right]
\end{eqnarray}

In the limit that $m_j\rightarrow \overline{m}$ and $m_r\rightarrow m_s$, these
expressions reduce down to those of QCD~\cite{J92}
with dynamical $\pi$'s, $K$'s and
$\eta$, but with the $\eta^\prime$ integrated out of the theory. 
In making this comparison, the leading order expressions for the meson masses,
$m_\pi^2 = 2 \overline{\lambda} \overline{m}$,
$m_K^2= \overline{\lambda} (\overline{m}+m_s)$, and 
$m_\eta^2 =  \overline{\lambda} {2\over 3}  (\overline{m}+2m_s)$,
have been used.
The contributions from graphs with intermediate states 
from the {\bf 138} dimensional representation (including the decuplet)
possess divergences proportional to $\Delta m_q$ and $\Delta^3$.
These divergences, and the associated renormalization scale dependence,
are absorbed by the counterterms $M_0$ and $\alpha_M$, $\beta_M$ and
$\sigma_M$.
The required scale dependences of the renormalized constants are
\begin{eqnarray}
M_0(\mu) & = & M_0-{5\over\overline{\lambda}} \Delta^2\Gamma
\ \ ,\ \ 
\alpha_M(\mu) \ =\ \alpha_M - {3\over 2}\Gamma\ \ ,
\nonumber\\
\beta_M(\mu) & = & \beta_M - {15\over 4}\Gamma
\ \ ,\ \ 
\sigma_M(\mu) \ =\ \sigma_M - {3\over 4}\Gamma
\ \ ,\ \ 
\Gamma \ =\  { {\cal C}^2\Delta\overline{\lambda}\over 12\pi^2 f^2}
\log(\mu)
\ \ \ ,
\end{eqnarray}
where we have used the leading order expressions for the meson masses.

\section{Magnetic Moments of the Octet-Baryons}

The magnetic moments of the octet-baryons have played a key role in the 
development of hadronic physics, and 
they have been studied with great success in QCD using 
$\chi$PT~\cite{JLMS92,MS97,DH98,PRM00,CP74}.
The expansion about the chiral limit takes the form
$\mu_B\ \sim\ \mu_0\ +\ \beta\  \sqrt{m_q}\ +\ \gamma\  m_q\log m_q\ 
+\ \delta\ m_q\ +\ ...$,
where each of the quantities $\mu_0, \beta, \gamma$, and $\delta$ have been
determined for each of the baryons~\cite{JLMS92,MS97,DH98,PRM00,CP74}.
Recently, the progress in quenched lattice simulations has lead to an
investigation of the chiral expansion of the magnetic moments
in Q$\chi$PT~\cite{S01a}.
Unlike QCD, the chiral limit of QQCD is found to be divergent due to 
terms of the form $\log m_q$ arising from hairpin interactions.
In this section we analyze the magnetic moments of the octet-baryons in 
PQ$\chi$PT including the leading one-loop contributions of the form
$\sim\sqrt{m_Q}$.

\subsection{Electric Charges in PQQCD}

An issue that was not addressed in  work of Ref.~\cite{S01a}
is  the non-uniqueness of the 
quark electric charge matrix in the sea- and ghost-sectors.
It was recently pointed out by Golterman and Pallante~\cite{GP01a}
that the flavor structure of QCD non-leptonic weak operators 
does not uniquely define the analogous non-leptonic weak operators 
in PQQCD.   Such an ambiguity is present for electromagnetic observables,
including the magnetic moments, and is also present for quantities
such as matrix elements of isovector twist-2 operators, that we discuss in the
next section.

In QCD the light quark electric charge matrix is 
${\cal Q}={\rm diag.}(+{2\over 3}, -{1\over 3}, -{1\over 3})$, 
which transforms as 
an ${\bf 8}$ under $SU(3)_V$, and one of the nice features of nature is
that there is no singlet component.
As there is a correspondence between operators in PQ$\chi$PT and 
$\chi$PT it is desirable that the electric charge matrix in PQQCD have
vanishing supertrace, so that additional operators are not introduced.
Therefore, the most general electric charge matrix that can be considered
is~\footnote{In principle, any supertraceless matrix can be used
to determine $\mu_D$ and $\mu_F$~\cite{MGpriv}, 
however, we restrict ourselves to matrices
that give matrix elements of ${\cal Q}$ in the $m_j\rightarrow \overline{m}$
and $m_r\rightarrow m_s$ limit.}
\begin{eqnarray}
{\cal Q}^{(PQ)} & = & 
{\rm diag}\left(\ +{2\over 3}\ ,\  -{1\over 3}\ ,\  -{1\over 3}
\ ,\ q_j\ ,\  q_l\ ,\  q_r\ ,\  q_j\ ,\  q_l\ ,\  q_r\ \right)
\ \ \ ,
\label{eq:PQcharge}
\end{eqnarray}
where the charge assignments in the sea- and ghost-sectors are correlated 
in order to recover QCD in the limit $m_j\rightarrow m_u$,
$m_l\rightarrow m_d$, and  $m_r\rightarrow m_s$.
The form of ${\cal Q}^{(PQ)}$ in eq.~(\ref{eq:PQcharge})
is automatically supertraceless.
While any choice of $q_j, q_l,$ and $q_r$ are as good as 
any other choice it is  useful to consider two cases.
First, when $q_j=+{2\over 3}$ and $q_l=q_r=-{1\over 3}$, an extension
of  the quenched operator used in  Ref.~\cite{S01a},
contributions from disconnected diagrams involving the valence- and 
ghost-quarks exactly cancel.  The only disconnected diagrams that contribute
are those involving the heavier sea-quarks (in the analogous quenched
calculation this corresponds to the absence of disconnected diagrams).
Second, $q_j=q_l=q_r=0$  corresponds to vanishing contributions from
disconnected diagrams involving the sea- and ghost-sectors but the presence of
disconnected diagrams involving the valence-quarks
(in the analogous quenched
calculation this corresponds to the presence of disconnected diagrams).
It is clear that different 
values of $q_j,q_l$ and $q_r$ correspond to different
weightings of the disconnected diagrams.

\subsection{The Magnetic Moments}

\begin{figure}[!ht]
\centerline{{\epsfxsize=4.0in \epsfbox{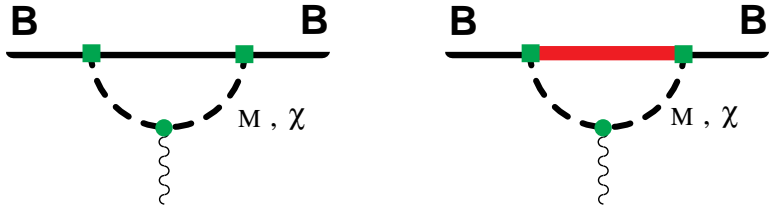}}} 
\vskip 0.15in
\noindent
\caption{\it 
One-loop graphs that give contributions of the form 
$\sim \sqrt{m_Q}$
to the magnetic moments of the octet-baryons.
A solid, thick-solid and dashed line denote an 
{\bf 240}-baryon, {\bf 138}-baryon, and a meson, respectively.
The solid-squares denote an axial coupling from eq.~(\ref{eq:ints})
}
\label{fig:magmoms}
\vskip .2in
\end{figure}

At leading order in the chiral expansion 
the magnetic moments of the octet-baryons
arise from two dimension-5 operators,
\begin{eqnarray}
{\cal L} & = & 
{e\over 4 M_N} F_{\mu\nu}\ 
\left[\ 
\mu_\alpha\ \left(\ \overline{\cal B}\ \sigma^{\mu\nu}\  {\cal B}\  
{\cal Q}_{\xi+}^{(PQ)}\ \right)
\ +\ 
\mu_\beta\ \left(\ \overline{\cal B}\ \sigma^{\mu\nu} \ 
{\cal Q}_{\xi+}^{(PQ)}\ 
{\cal B}\ \right)
\ \right]
\ \ \ ,
\label{eq:dimfive}
\end{eqnarray}
where 
${\cal Q}_{\xi+}^{(PQ)}  =  {1\over 2} 
\left(\ \xi^\dagger {\cal Q}^{(PQ)}\xi + \xi {\cal Q}^{(PQ)}\xi^\dagger
\right)$,
and where the index contractions are defined in eq.~(\ref{eq:Contractions}).
By considering only the ground floor baryons we can make the identification
with the flavor structure of operators commonly used in $\chi$PT
\begin{eqnarray}
{\cal L} & = & 
{e\over 4 M_N} F_{\mu\nu}\ 
\left(\ 
\mu_D\ {\rm Tr}\left[\ \overline{B}\sigma^{\mu\nu} 
\{ {\cal Q}_{\xi_+} ,B\}\ \right]
\ +\ 
\mu_F\ {\rm Tr}\left[\ \overline{B}\sigma^{\mu\nu} 
\left[{\cal Q}_{\xi_+},B\right]\ \right]
\ 
\right)
\ +\ ...
\ \ \ ,
\label{eq:magQCD}
\end{eqnarray}
where the ellipses denotes terms involving the meson field,
and find that
\begin{eqnarray} 
\mu_\alpha & = & {2\over 3}\mu_D+2\mu_F
\ \ ,\ \ 
\mu_\beta \ = \ -{5\over 3}\mu_D + \mu_F
\ \ \ .
\end{eqnarray}
The NLO contribution to the magnetic moments is of the form $\sqrt{m_Q}$,
arising from the one-loop diagrams shown in  Fig.~\ref{fig:magmoms}.
Up to this order, we write the magnetic moment of the $i$-th baryon as
\begin{eqnarray}
\mu_i & = & \alpha_i\ +\ {M_N\over 4\pi f^2}\ \left[\ 
\beta_i\ +\ \beta_i^\prime\ \right]
\ \ \ ,
\label{eq:magdef}
\end{eqnarray}
where $\alpha_i$ is the tree-level contribution, $\beta_i$ is the contribution
from the diagrams shown in Fig.~\ref{fig:magmoms} with baryons in the 
{\bf 240} representation (including the octet) 
in the intermediate state, and 
$\beta_i^\prime$ is the contribution
from the diagrams shown in Fig.~\ref{fig:magmoms} with baryons in the 
{\bf 138} representation (including the decuplet)
in the intermediate state.

Explicit computation of the one-loop diagrams in 
Fig.~\ref{fig:magmoms} give contributions to the proton magnetic moment
\begin{eqnarray}
\alpha_p & = & {1\over 3}\ \mu_D\ +\ \mu_F
\nonumber\\
\beta_p & = & 
-{1\over 9}\left(7D^2+6DF-9F^2\right) m_\pi
-{1\over 9}\left(5D^2-6DF+9F^2\right) m_K
\nonumber\\
& & -{1\over 9}\left(D+3F\right)^2 \left(2m_{ju}+m_{ru}\right)
\nonumber\\
& &  + \ {1\over 3}\left( 5D^2-6DF+9F^2\right)
\left[\ q_{jl}\left(m_{ju}-m_\pi\right) + q_r\left(m_{ru}-m_K\right) 
\ \right]
\nonumber\\
\beta^\prime_p & = & {{\cal C}^2\over 6}\ 
\left[\ 
-{4\over 3}\cf_\pi+{1\over 3}\cf_K
- q_{jl}\left(\cf_{ju}-\cf_\pi\right) - q_r\left(\cf_{ru}-\cf_K\right)
\right]
\ \ \ ,
\label{eq:pmag}
\end{eqnarray}
where $q_{jl} = q_j+q_l$.
The function arising from the decuplet loops is the 
standard one for magnetic moments~\cite{JLMS92}, 
and we use the shorthand notation
${\cal F}_i={\cal F}(m_i,\Delta,\mu)$, with
\begin{eqnarray}
\pi {\cal F}(m,\Delta,\mu)
& = & \sqrt{\Delta^2-m^2}\log\left({\Delta-\sqrt{\Delta^2-m^2+i\epsilon}
\over \Delta+\sqrt{\Delta^2-m^2+i\epsilon}}\right)
\ -\ \Delta\log\left({m^2\over\mu^2}\right)
\ \ \ .
\label{eq:magfun}
\end{eqnarray}

For the neutron we find
\begin{eqnarray}
\alpha_n & = & -{2\over 3}\ \mu_D
\nonumber\\
\beta_n & = & 
{1\over 9}\left(17D^2-6DF+9F^2\right) m_\pi
-{1\over 9}\left(5D^2-6DF+9F^2\right) m_K
\nonumber\\
& & -{4\over 9} D(D-3F)\left(2m_{ju}+m_{ru}\right)
\nonumber\\
& &  + \ {1\over 3}\left( 5D^2-6DF+9F^2\right)
\left[\ q_{jl}\left(m_{ju}-m_\pi\right) + q_r\left(m_{ru}-m_K\right) 
\ \right]
\nonumber\\
\beta^\prime_n & = & {{\cal C}^2\over 18}\ 
\left[\ 
2\cf_\pi + \cf_K+ 2 \cf_{ju} + \cf_{ru}
- 3 q_{jl}\left(\cf_{ju}-\cf_\pi\right) - 3 q_r\left(\cf_{ru}-\cf_K\right)
\right]
\ \ \ ,
\label{eq:nmag}
\end{eqnarray}

For the $\Sigma^+$ we find
\begin{eqnarray}
\alpha_{\Sigma^+} & = & {1\over 3}\ \mu_D\ +\ \mu_F
\nonumber\\
\beta_{\Sigma^+} & = &
{2\over 9}\left(D^2+3F^2\right)\left[\ m_\pi-4 m_{ju}-2 m_{ru}\ \right]
-{1\over 3}\left(D-F\right)^2\left[\ m_{\eta_s}-2 m_{js} - m_{rs}\right]
\nonumber\\ & & 
-{1\over 9}\left( 11D^2 + 6 DF + 3 F^2\right) m_K
+{2\over 3}\left(D^2+3F^2\right) 
\left[ q_{jl} (m_{ju}-m_\pi) + q_r (m_{ru}-m_K)\right]
\nonumber\\ & & 
+\left(D-F\right)^2
\left[ q_{jl} (m_{js}-m_K) + q_r (m_{rs}-m_{\eta_s})\right]
\nonumber\\
\beta_{\Sigma^+}^\prime & = & 
{{\cal C}^2\over 54}\ \left[\ 
-\cf_\pi - 10\cf_K+ 2 \cf_{\eta_s} + 4 \cf_{ju} + 2 \cf_{ru}
-4\cf_{js} - 2 \cf_{rs}
\right. \nonumber\\
& & \left.
- 3 q_{jl}\left[ \cf_{ju}-\cf_\pi + 2 \cf_{js} - 2 \cf_K \right]
- 3 q_r\left[ \cf_{ru}-\cf_K + 2 \cf_{rs} - 2 \cf_{\eta_s}\right]
\ \right]
\ \ \ ,
\label{eq:spmag}
\end{eqnarray}
and for the $\Sigma^-$ we find
\begin{eqnarray}
\alpha_{\Sigma^-} & = & {1\over 3}\ \mu_D\ -\ \mu_F
\nonumber\\
\beta_{\Sigma^-} & = & 
{2\over 9}\left(D^2+3F^2\right)\left[\ m_\pi+2 m_{ju}+ m_{ru}\ \right]
-{1\over 3}\left(D-F\right)^2\left[\ m_{\eta_s}-2 m_{js} - m_{rs}\right]
\nonumber\\ & & 
+{1\over 9}\left( D^2 - 6 DF - 3 F^2\right) m_K
+{2\over 3}\left(D^2+3F^2\right) 
\left[ q_{jl} (m_{ju}-m_\pi) + q_r (m_{ru}-m_K)\right]
\nonumber\\ & & 
+\left(D-F\right)^2
\left[ q_{jl} (m_{js}-m_K) + q_r (m_{rs}-m_{\eta_s})\right]
\nonumber\\
\beta_{\Sigma^-}^\prime & = &
{{\cal C}^2\over 54}\left[\
-\cf_\pi - \cf_K+ 2 \cf_{\eta_s} -2 \cf_{ju} - \cf_{ru}
-4\cf_{js} - 2 \cf_{rs}
\right. \nonumber\\
& & \left.
- 3 q_{jl}\left[ \cf_{ju}-\cf_\pi + 2 \cf_{js} - 2 \cf_K \right]
- 3 q_r\left[ \cf_{ru}-\cf_K + 2 \cf_{rs} - 2 \cf_{\eta_s}\right]
\ \right]
\ \ \ .
\label{eq:smmag}
\end{eqnarray}

For the $\Lambda$ we find
\begin{eqnarray}
\alpha_{\Lambda} & = & -{1\over 3}\ \mu_D
\nonumber\\
\beta_{\Lambda} & = & 
{1\over 27}\left(D+3F\right)^2 \left[\ 2 m_{js}-m_{\eta_s} +m_{rs}\ \right]
-{1\over 27}\left( 7D^2-12 DF + 9F^2\right) 
\left[ m_{ru}- 2 m_\pi + 2 m_{ju}\right]
\nonumber\\ & & 
+{1\over 27} \left( 5D^2+30 DF - 9F^2\right) m_K
+{1\over 9}\left(D+3F\right)^2 
\left[\ q_{jl}\left( m_{js}-m_K\right) + q_r\left(m_{rs}-m_{\eta_s}\right)
\ \right]
\nonumber\\ & & 
+{2\over 9} \left( 7D^2-12 DF + 9F^2\right) 
\left[\ q_{jl}\left( m_{ju}-m_\pi\right) + q_r\left(m_{ru}-m_K\right)
\ \right]
\nonumber\\
\beta_\Lambda^\prime & = & 
{{\cal C}^2\over 36} \left[\ 
2 \cf_{ju}-2\cf_\pi+5\cf_K+\cf_{ru}
-6 q_{jl}\left(\cf_{ju}-\cf_\pi\right) 
-6 q_r\left(\cf_{ru}-\cf_K\right) 
\ \right]
\ \ \ .
\label{eq:Lmag}
\end{eqnarray}

The $\Lambda-\Sigma^0$ transition is somewhat special in that it does not
depend upon the $q_i$'s,
\begin{eqnarray}
3\sqrt{3}\ \alpha_{\Lambda\Sigma} & = & 3 \mu_D
\nonumber\\
3\sqrt{3}\ \beta_{\Lambda\Sigma} & = &
-\ 2 D^2 (2 m_\pi + m_K)\ +\ 2 D(D-3F) (m_{ru}+2 m_{ju})
\nonumber\\
3\sqrt{3}\ \beta_{\Lambda\Sigma}^\prime & = &
-{{\cal C}^2\over 4}\left[\ 2 {\cal F}_\pi + {\cal F}_K +  2 {\cal F}_{ju}
 + {\cal F}_{ru}\ \right]
\ \ \ .
\label{eq:LSmag}
\end{eqnarray}

For the $\Xi^0$ we find
\begin{eqnarray}
\alpha_{\Xi^0} & = & -{2\over 3}\ \mu_D
\nonumber\\
\beta_{\Xi^0} & = & 
{2\over 9}\left(D^2+3F^2\right) \left[\ m_{rs}-m_{\eta_s}+2 m_{js}\ \right]
+{1\over 3}\left(D-F\right)^2 \left[\ m_\pi - 2 m_{ru} - 4 m_{ju}\ \right]
\nonumber\\ & & 
+{1\over 9}\left( 11D^2 +6 DF+3F^2\right) m_K
+ {2\over 3}\left(D^2+3F^2\right) \left[\ q_{jl}\left(m_{js}-m_K\right)
+ q_r \left(m_{rs}-m_{\eta_s}\right) \right]
\nonumber\\ & & 
+\left(D-F\right)^2 \left[\ q_{jl}\left(m_{ju}-m_\pi\right)
+ q_r \left(m_{ru}-m_K\right) \right]
\nonumber\\
\beta_{\Xi^0}^\prime & = & 
{{\cal C}^2\over 54}\left[\ 
10 \cf_K - 2 \cf_\pi +\cf_{\eta_s} + 8 \cf_{ju} + 4 \cf_{ru}
-2 \cf_{js} - \cf_{rs}
\right.\nonumber\\ & & \left.\qquad
- 3 q_{jl}\left(2\cf_{ju}-2\cf_\pi+\cf_{js}-\cf_K\right)
-3 q_r \left(2\cf_{ru}-2\cf_K+\cf_{rs}-\cf_{\eta_s}\right)
\ \right]
\ \ \ ,
\label{eq:C0mag}
\end{eqnarray}
and finally, for the $\Xi^-$ we find
\begin{eqnarray}
\alpha_{\Xi^-} & = & {1\over 3}\ \mu_D\ -\ \mu_F
\nonumber\\
\beta_{\Xi^-} & = & 
{1\over 3}\left(D-F\right)^2\left[\ m_\pi+2 m_{ju}+m_{ru}\ \right]
+{2\over 9}\left(D^2+3F^2\right)\left[\  2m_{js}+m_{rs}-m_{\eta_s}\ \right]
\nonumber\\ & & 
-{1\over 9}\left( D^2-6 DF -3 F^2\right) m_K
+{2\over 3}\left(D^2+3F^2\right)\left[\ q_{jl} \left(m_{js}-m_K\right)
+q_r \left(m_{rs}-m_{\eta_s}\right)\right]
\nonumber\\ & & 
+\left(D-F\right)^2\left[\  q_{jl} \left(m_{ju}-m_\pi\right)
+q_r \left(m_{ru}-m_K\right)\right]
\nonumber\\
\beta_{\Xi^-}^\prime & = & 
{{\cal C}^2\over 54}\left[\ 
\cf_K-2\cf_\pi+\cf_{\eta_s} -4\cf_{ju}-2\cf_{ru}-2\cf_{js}-\cf_{rs}
\right.\nonumber\\ & & \left.
-3 q_{jl}\left(2\cf_{ju}-2\cf_\pi+\cf_{js}-\cf_K\right)
-3 q_r\left(2\cf_{ru}-2\cf_K+\cf_{rs}-\cf_{\eta_s}\right)
\ \right]
\ \ \ .
\label{eq:Cmmag}
\end{eqnarray}

Each of the expressions for $\alpha_i$, $\beta_i$ and $\beta_i^\prime$
reduce down to those of QCD when $m_j\rightarrow\overline{m}$ and 
$m_r\rightarrow m_s$, independent of the choice of the $q_i$, as expected.
The divergences and renormalization scale dependence
associated with the ${\cal F}_i$ functions can be removed by defining
\begin{eqnarray}
\mu_D (\mu) &  = & 
\mu_D\ +\ {{\cal C}^2 M_N \Delta\over 4\pi^2 f^2} \log \mu
\ \ \ ,\ \ \ \mu_F (\mu) \ =\ \mu_F
\ \ \ ,
\end{eqnarray}
independent of the charges $q_i$.

For arbitrary choices of the $q_i$ and quark-masses the 
Caldi-Pagels relations~\cite{CP74}
between the magnetic moments are not satisfied.
However, the relation found in Ref.~\cite{JLMS92}
\begin{eqnarray}
6\ \mu_\Lambda\ +\ \mu_{\Sigma^-}\ -\ 4\sqrt{3}\ \mu_{\Lambda\Sigma}
\ & = & 
\ 4\ \mu_n\ -\ \mu_{\Sigma^+}\ +\ 4\ \mu_{\Xi^0}
\ \ \ ,
\end{eqnarray}
that is valid up to order $m_q\log m_q$ in $\chi$PT 
is found to hold in PQ$\chi$PT up to order $\sqrt{m_q}$.
We have not performed the expansion to higher orders in 
PQ$\chi$PT to determine if the relation persists at higher orders.

It is important to emphasize that in the limit where the sea-quark masses
become equal to the valence quark masses, the magnetic moments computed with 
PQ$\chi$PT are those of QCD for any value of the charges $q_i$.
Therefore, the value of the $m_Q$-independent counterterms, 
$\mu_D$ and $\mu_F$, that are determined by lattice simulations
and eq.~(\ref{eq:magdef})
should be independent of the choice of
sea- and ghost-quark electric charges $q_i$ in eq.~(\ref{eq:PQcharge}). 
This is, of course,  modulo contributions from higher
orders in the chiral expansion.
It is conceivable that there is a choice of the $q_i$'s, 
corresponding
to an optimal weighting of the disconnected diagrams that minimizes the 
uncertainty in the determination of $\mu_D$ and $\mu_F$.
To illustrate this point, consider the magnetic moment of the proton.
If one chooses
\begin{eqnarray}
q_{jl} & = & { 2 (D+3F)^2 \over 3 (5D^2-6 DF + 9 F^2)}
\ \ {\rm and }\ \ 
q_{r} \ = \ {  (D+3F)^2 \over 3 (5D^2-6 DF + 9 F^2)}
\ \ \ ,
\end{eqnarray}
the one loop contributions proportional to $m_{ju}$ and
$m_{ru}$ from diagrams with intermediate state baryons in
the  {\bf 240} dimensional representation vanish, 
leaving contributions dependent upon the sea-quark masses only
from diagrams with intermediate state baryons in the {\bf 138} dimensional 
representation.
It is clear that one can determine numerical values for the 
$q_{jl}$ and $q_r$ that minimize the one-loop dependence upon the 
sea-quark masses.
Unfortunately, the optimal choice of $q_{jl}$ and $q_r$ will be different for
each member of the octet.
However, it may well be the case that examining the behavior of the chiral
extrapolation for many different values of the $q_i$ may yield valuable
information about, not only $\mu_D$ and $\mu_F$, but the convergence of
the expansion itself.

\section{Forward Matrix Elements of Isovector Twist-2 Operators}

The forward matrix elements of twist-2 operators
play an important role in hadronic structure, as they are directly related
to the moments of the parton distribution functions.
Recently, it was realized that the long-distance contributions to 
these matrix elements could be computed
order-by-order in the chiral
expansion using chiral perturbation theory~\cite{AS,CJ,CJb}.
These corrections have been applied to results from both quenched and 
unquenched lattice data~\cite{aussies}.
In addition, the long-distance contributions arising in QQCD
and the large-$N_c$ limit of QCD
have been computed in Ref.~\cite{CSqqcd} and Ref.~\cite{CJNc}, 
respectively.
Further, this technique has been applied to the off-forward
matrix elements of twist-2 operators in order 
to study the spin structure of the  proton~\cite{CJoff}.

In QCD, the nonsinglet operators have the form,
\begin{eqnarray}
{\cal O}^{ (n), a}_{\mu_1\mu_2\ ... \mu_n}
& = & 
{1\over n!}\ 
\overline{q}\ \lambda^a\ \gamma_{ \{\mu_1  } 
\left(i \stackrel{\leftrightarrow}{D}_{\mu_2}\right)\ 
... 
\left(i \stackrel{\leftrightarrow}{D}_{ \mu_n\} }\right)\ q
\ -\ {\rm traces}
\ \ \ ,
\label{eq:twistop}
\end{eqnarray}
where the $\{ ... \}$ denotes symmetrization on all Lorentz indices,
and where $\lambda^a$ are Gell-Mann matrices acting in flavor-space.
They transform as $({\bf 8},{\bf 1})\oplus  ({\bf 1},{\bf 8})$
under $SU(3)_L\otimes SU(3)_R$ chiral transformations~\cite{AS,CJ}.
Of particular interest to us are the isovector operators where
$\lambda^3={\rm diag}(1,-1,0)$.

In PQQCD the nonsinglet operators have the form
\begin{eqnarray}
^{PQ}{\cal O}^{(n), a}_{\mu_1\mu_2\ ... \mu_n}
& = & 
{1\over n!}\ 
\overline{Q}\ \overline{\lambda}^a\ \gamma_{ \{\mu_1  } 
\left(i \stackrel{\leftrightarrow}{D}_{\mu_2}\right)\ 
... 
\left(i \stackrel{\leftrightarrow}{D}_{ \mu_n\} }\right)\ Q
\ -\ {\rm traces}
\ \ \ ,
\label{eq:Qtwistop}
\end{eqnarray}
where the $\overline{\lambda}^a$ are super Gell-Mann matrices,
and the same ambiguity exists in extending the $\lambda^3$ 
in QCD to $\overline{\lambda}^3$ in PQQCD.
With the requirement that $\overline{\lambda}^3$ is supertraceless
and QCD is recovered in the limit $m_j\rightarrow \overline{m}$ and 
$m_r\rightarrow m_s$, the most general  flavor structure 
for $\overline{\lambda}^3$ is
\begin{eqnarray}
\overline{\lambda}^3 & = & 
\left(\ 1\  ,\  -1\ ,\  0 \ ,\  y_j \ ,\  y_l \ ,\  
y_r \ ,\   y_j \ ,\  y_l \ ,\  y_r\ \right)
\ \ \ .
\label{eq:isocharge}
\end{eqnarray}
For an arbitrary choice of the $y_i$, this operator contains both 
isovector and isoscalar components.
It is purely isovector only when $y_j+y_l=0$ and $y_r=0$.
As result, for arbitrary $y_i$, the usual isovector relations between
matrix elements do not hold. 
The fact that disconnected diagrams can only be isoscalar makes this result
obvious.

At leading order in the chiral expansion, matrix elements of the isovector
operator
$^{PQ}{\cal O}^{(n), 3}_{\mu_1\mu_2\ ... \mu_n}$
are reproduced by operators of the form~\cite{AS}
\begin{eqnarray}
^{PQ}{\cal O}^{(n),3}_{\mu_1\mu_2 ...\mu_n}
& & \rightarrow 
a^{(n)} \left(i\right)^n {f^2\over 4} 
\left({1\over\Lambda_\chi}\right)^{n-1}
{\rm str}\left[\ 
\Sigma^\dagger  \overline{\lambda}^3 \overrightarrow\partial_{\mu_1}
 \overrightarrow\partial_{\mu_2}...
 \overrightarrow\partial_{\mu_n}
\Sigma
\ +\ 
\Sigma \overline{\lambda}^3 \overrightarrow\partial_{\mu_1}
 \overrightarrow\partial_{\mu_2}...
 \overrightarrow\partial_{\mu_n}
\Sigma^\dagger \right]
\nonumber\\
& +  & \alpha^{(n)}\ v_{\mu_1} v_{\mu_2}...v_{\mu_n}\ 
\left(\ \overline{\cal B}\  {\cal B}\  \overline{\lambda}^3_{\xi +}\ \right)
\ +\ 
\beta^{(n)}\ v_{\mu_1} v_{\mu_2}...v_{\mu_n}\ 
\left(\ \overline{\cal B}\  \overline{\lambda}^3_{\xi +}\  {\cal B}\ \right)
\nonumber\\
& + &   
\gamma^{(n)} 
\ v_{\mu_1} v_{\mu_2}...v_{\mu_n}\ 
\left(\ \overline{\cal T}^\alpha\  \overline{\lambda}^3_{\xi +}
\ {\cal T}_\alpha
\right)
\ +\ 
\sigma^{(n)} {1\over n !}
\ v_{\{ \mu_1} v_{\mu_2}...v_{\mu_{n-2}}\ 
\left(\ \overline{\cal T}_{\mu_{n-1}}\  \overline{\lambda}^3_{\xi +}\ 
{\cal T}_{\mu_n\}}
\right)
\nonumber\\ & & 
\ -\ {\rm traces}
\ \ \ .
\label{eq:tree}
\end{eqnarray}
In general, the coefficients $a^{(n)}, \alpha^{(n)},
\beta^{(n)}, \gamma^{(n)}$ and
$\sigma^{(n)}$ are not constrained by symmetries and
must be determined from elsewhere. However for $n=1$ they are
fixed by the isospin charge of the hadrons to be 
\begin{eqnarray}
a^{(1)} & = & +1
\ \ \ ,\ \ \ 
\alpha^{(1)} \ = \ +2
\ \ \ ,\ \ \ 
\beta^{(1)}\ =\ +1
\ \ \ ,\ \ \ 
\gamma^{(1)}\ =\ -3
\ \ \ ,\ \ \ 
\sigma^{(1)}\ =\ 0
\ \ \ .
\end{eqnarray}

At NLO there are contributions from counterterms involving one 
insertion of the quark mass matrix $m_Q$,
\begin{eqnarray}
^{PQ}{\cal O}^{(n),3}_{\mu_1\mu_2 ...\mu_n}
& &\rightarrow 
\left[\ 
b_1\  \cbb^{kji}\ \{\  \overline{\lambda}^3_{\xi +}\ ,\ 
{\cal M}_+\ \}^n_i\ \cb_{njk}
\right.\nonumber\\ & & \left.
+\ 
b_2\ (-)^{(\eta_i+\eta_j)(\eta_k+\eta_n)}\ 
\cbb^{kji}\ \{\  \overline{\lambda}^3_{\xi +}\ ,\ {\cal M}_+\ \}^n_k\ \cb_{ijn}
\right.\nonumber\\ & & \left.
+\ 
b_3\  (-)^{\eta_l (\eta_j+\eta_n)}\
\cbb^{kji}\  \left(\overline{\lambda}^3_{\xi +}\right)^l_i\ 
\left( {\cal M}_+\right)^n_j
\cb_{lnk}
\right.\nonumber\\ & & \left.
+\ 
b_4 \  (-)^{\eta_l \eta_j}\ 
\cbb^{kji}\ \left(  
\left(\overline{\lambda}^3_{\xi +}\right)^l_i\ \left( {\cal M}_+\right)^n_j
\ +\ \left( {\cal M}_+\right)^l_i \left(\overline{\lambda}^3\right)^n_j \right)
\cb_{nlk}
\right.\nonumber\\ & & \left.
+\ b_5\  (-)^{\eta_i(\eta_l+\eta_j)}\ 
\cbb^{kji} \left(\overline{\lambda}^3_{\xi +}\right)^l_j 
\left( {\cal M}_+\right)^n_i
\cb_{nlk}
\ +\ b_6\  \cbb^{kji}  \left(\overline{\lambda}^3_{\xi +}\right)^l_i \cb_{ljk}
\ {\rm str}\left( {\cal M}_+ \right) 
\right.\nonumber\\ & & \left.
\ +\ b_7\  \ (-)^{(\eta_i+\eta_j)(\eta_k+\eta_n)}\ 
\cbb^{kji}  \left(\overline{\lambda}^3_{\xi +}\right)^n_k \cb_{ijn}
\ {\rm str}\left( {\cal M}_+ \right) 
\right.\nonumber\\ & & \left.
+\ b_8\ \cbb^{kji}\ \cb_{ijk} 
\ {\rm str}\left(\overline{\lambda}^3_{\xi +}\   {\cal M}_+ \right) 
\right.\nonumber\\ & & \left.
\right]\ v_{\mu_1} v_{\mu_2}...v_{\mu_n}\ 
\ -\ {\rm traces}
\ ,
\label{eq:tcts}
\end{eqnarray}
where the coefficients $b_1,...b_8$ are to be determined.
We find that there are eight counterterms, one more than in 
ordinary $SU(3)$.
It is interesting to note that in restricting oneself to external states 
that involve only octet-baryons there are contributions from
eq.~(\ref{eq:tcts}) that are of the form $m_u+m_d+m_s$ and $m_j+m_l+m_r$.
The former arises from the trace of the valence-quark mass matrix that arises
from the $SU(3)$ contractions when the indices in $B_{ijk}$ are restricted to 
$1,2,3$, 
while the later arises from the supertrace of the full mass matrix $m_Q$.

\begin{figure}[!ht]
\centerline{{\epsfxsize=3.0in \epsfbox{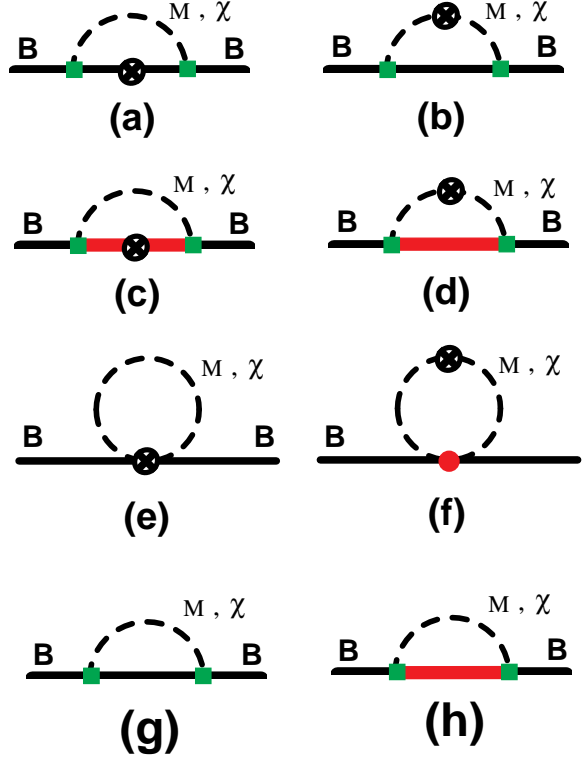}}} 
\vskip 0.15in
\noindent
\caption{\it 
One-loop graphs that give contributions of the form 
$\sim m_Q \log m_Q$
to the matrix elements of the isovector twist-2 operators
in the octet-baryons.
A solid, thick-solid and dashed line denote an 
{\bf 240}-baryon, {\bf 138}-baryon, and a meson, respectively.
The solid-squares denote an axial coupling given in eq.(\ref{eq:ints}),
while the solid circle denotes an insertion of the strong two-pion-nucleon
interaction given in eq.(\ref{eq:free}).
The crossed circle denotes an insertion of the tree-level matrix element
of $^{PQ}{\cal O}^{(n), a}_{\mu_1\mu_2\ ... \mu_n}$.
Diagrams (a) to (f) are vertex corrections, while 
diagrams (g) and (h) give rise to wavefunction renormalization.
}
\label{fig:twist}
\vskip .2in
\end{figure}

At one-loop level the forward matrix elements of 
$^{PQ}{\cal O}^{(n),3}_{\mu_1\mu_2 ...\mu_n}$
in the $i$-th baryon can be  written as
\begin{eqnarray}
\langle ^{PQ}{\cal O}^{(n),3}_{\mu_1\mu_2 ...\mu_n} \rangle_i
& = & v_{\mu_1} v_{\mu_2}...v_{\mu_n}\ 
\left[\ 
\rho_i^{(n)}
\ +\ {1-\delta^{n1}\over 16\pi^2 f^2}
\left(\ 
\eta_i^{(n), 0}\ -\ \rho_i^{(n)} w_i\ +\ y_{jl}\ \eta^{(n), j}_i
\ +\ y_r\ \eta^{(n), r}_i
\ \right)
\right.\nonumber\\ & & \left.\qquad\qquad\qquad\qquad
c_i^{(n), 0} +\ y_{jl}\ c^{(n), j}_i
\ +\ y_r\ c^{(n), r}_i
\ \right]
\ \ -\ \ {\rm traces}
\ \ \ ,
\label{eq:ttmat}
\end{eqnarray}
where $\rho_i^{(n)}$ are the tree-level contributions.
The factor of $1-\delta^{n1}$ appears in the higher order corrections
because the isovector charge is not renormalized.
The diagrams shown in Fig.\ref{fig:twist} give the leading non-analytic
contributions to
the wavefunction renormalization,$w_i$,
the vertex contributions, $\eta_i^{(n), 0}$,
that are independent of
the charges of the ghost- and sea-quarks, and to the vertex contributions,
$\eta_i^{(n), j,l,r}$, are contributions associated with the 
charges of the ghost- and valence quarks.
The contributions proportional to  $y_j$ and $y_l$ are equal 
due to the 
sea-quark isospin symmetry of the theory,
which can be seen straightforwardly by considering
quark-line diagrams. 
Therefore, we combine the contributions into one by 
defining $y_{jl} = y_j\ +\ y_l$.

For the proton matrix element we find
\begin{eqnarray}
\rho_p^{(n)} & = & 
{1\over 3}\  \left(2 \alpha^{(n)}-\beta^{(n)}\right) 
\nonumber\\
w_p & = & -4D(D-3F) L_\pi\ +\ (5D^2-6DF+9F^2)\ (2L_{ju}+L_{ru})
\ +\ 3 (D-3F)^2\ R_{\pi , \pi}
\nonumber\\
& & 
+\ {\cal C}^2\ \left( 2 J_\pi + 2 J_{ju} + J_{ru}\ \right)
\nonumber\\
\eta_p^{(n), 0} & = & 4 D (F-D) (\ \alpha^{(n)} +\beta^{(n)}\ ) L_\pi 
 +\ 3\rho_p^{(n)} (D-3F)^2\ R_{\pi , \pi} 
\nonumber\\
& & +\ \left(\ 2(D^2- 3 DF + 2 F^2)\alpha^{(n)} + 2(D^2-F^2) \beta^{(n)}
\ -\ \rho_p^{(n)}\ \right)
\left(\ 2 L_{ju} + L_{ru}\ \right)
\nonumber\\
& & -\ 
{4 {\cal C}^2 \over 9}\ \left(\gamma^{(n)}-{\sigma^{(n)}\over 3}\right)
\left(\ 3 J_\pi\ + 2 J_{ju} + J_{ru}\ \right)
\nonumber\\
\eta_p^{(n), j} & = &
\left(L_{ju}-L_\pi\right)
\left((D^2+3 F^2)\alpha^{(n)} + 3 (D-F)^2\beta^{(n)}
+\alpha^{(n)}+\beta^{(n)}\right)
\nonumber\\ & & 
\ -\  \left(J_{ju}-J_\pi\right)
{{\cal C}^2 \over 3}\left(\gamma^{(n)}-{\sigma^{(n)}\over 3}\right)
\nonumber\\
\eta_p^{(n), r} & = &   \left(L_{ru}-L_K\right)
\left((D^2+3F^2)\alpha^{(n)} + 3 (D-F)^2 \beta^{(n)}
+\alpha^{(n)}+\beta^{(n)}\right)
\nonumber\\ & & 
\ -\  \left(J_{ru}-J_K\right)
{{\cal C}^2 \over 3}\left(\gamma^{(n)}-{\sigma^{(n)}\over 3}\right)
\nonumber\\
c_p^{(n), 0} & = & 
{1\over 3}\overline{m}\left(-2 b_1 + 4 b_2 - b_3 + b_4 + 2 b_5\right) 
+ {1\over 3}(2 b_7-b_6)\left(2 m_j+m_r\right)
\nonumber\\
c_p^{(n), j} & = & b_8 (m_j -\overline{m}) 
\nonumber\\
c_p^{(n), r} & = & b_8 (m_r-m_s) 
\ \ \ .
\label{eq:tp}
\end{eqnarray}
We have used the short-hand notation $L_\pi =
m_\pi^2\log\left({m_\pi^2/\mu^2}\right)$, and similarly for the other
$L_x$'s. 
The function arising from loops involving the {\bf 138} representation
is denoted by $J_a~=~J(m_a, \Delta, \mu)$~\cite{AS,CJ},
\begin{eqnarray}
J(m,\Delta,\mu) & = & 
\left(m^2-2\Delta^2\right)\log\left({m^2\over\mu^2}\right)
+2\Delta\sqrt{\Delta^2-m^2}
\log\left({\Delta-\sqrt{\Delta^2-m^2+ i \epsilon}\over
\Delta+\sqrt{\Delta^2-m^2+ i \epsilon}}\right)
\ \ \ .
\label{eq:decfun}
\end{eqnarray}
Also, we have defined the function
$R_{\pi , \pi} = {\cal H}(L_\pi, L_\pi, L_X)$.
The neutron matrix elements can be determined from the 
proton matrix elements by the replacements
\begin{eqnarray}
w_n & = & w_p
\ \ ,\ \ 
\eta_n^{(n), j}\ =\ +\eta_p^{(n), j}
\ \ ,\ \ 
\eta_n^{(n), r}\ =\ +\eta_p^{(n), r}
\ \ ,\ \ 
c_n^{(n), j}\ =\ +c_p^{(n), j}
\ \ ,\ \ 
c_n^{(n), r}\ =\ +c_p^{(n), r}
\nonumber\\
\rho_{n}^{(n)} & = & -\rho_{p}^{(n)}
\ \ ,\ \ 
\eta_n^{(n), 0}\ =\ -\eta_p^{(n), 0}
\ \ ,\ \ 
c_n^{(n), 0}\ =\ -c_p^{(n), 0}
\ \ \ .
\label{eq:tn}
\end{eqnarray}

For the $\Sigma^+$ we find
\begin{eqnarray}
\rho_{\Sigma^+}^{(n)} & = & {1\over 6}\  \left(5 \an+2\bn\right) 
\nonumber\\
w_{\Sigma^+} & = &  2(3F^2-D^2) L_\pi - 2 (D^2-6DF+3F^2) L_K 
\nonumber\\ & & 
+ 2(D^2+3F^2)(2 L_{ju}+L_{ru}) + 3 (D-F)^2 (2 L_{js}+L_{rs})
\nonumber\\
& & + 3 \left[\ 4 F^2 R_{\pi , \pi} + 4 F(F-D) R_{\pi , \eta_s}
+ (D-F)^2 R_{\eta_s , \eta_s}\ \right]
\nonumber\\
& & +\ {{\cal C}^2\over 3} \left[\ J_\pi + 5 J_K+2 J_{ju}+J_{ru}
+4 J_{js} + 2 J_{rs}\ \right]
 +\ {2{\cal C}^2\over 3}  
\left[\ {\cal T}_{\pi , \pi} +  {\cal T}_{\eta_s , \eta_s}
-  2 {\cal T}_{\pi , \eta_s}\ \right]
\nonumber\\
\eta_{\Sigma^+}^{(n), 0} & = & 
\left(\ F^2(5\an+2\bn)-D^2(\an+2\bn)\ \right) L_\pi
\nonumber\\
& & -\left(\ (5F^2-8DF+3D^2)\an + 2(F^2-4DF+D^2)\bn\ \right) L_K
\nonumber\\
& & +\ \left(\ {1\over 2}(5F^2-2DF+D^2)\an + (D+F)^2\bn\ -
\rho_{\Sigma^+}^{(n)} \right)
\left(2 L_{ju}+L_{ru}\right)
\nonumber\\
& & +\ {1\over 2}(D-F)^2 (5\an+2\bn) \left( 2 L_{js}+L_{rs}\right)
\nonumber\\
& & + 3 \rho_{\Sigma^+}^{(n)}
\left[\ 4 F^2 R_{\pi , \pi} + 4 F(F-D) R_{\pi , \eta_s}
+ (D-F)^2 R_{\eta_s , \eta_s}\ \right]
\nonumber\\
& & -\ {{\cal C}^2\over 9} \left(\gn-{\sn\over 3}\right)
\left[\ 2 J_\pi + 13 J_K + 2 J_{ju}+J_{ru}+8J_{js}+4 J_{rs}\ \right]
\nonumber\\
& & -\ {4{\cal C}^2\over 9} \left(\gn-{\sn\over 3}\right)
\left[\ {\cal T}_{\pi , \pi} +  {\cal T}_{\eta_s , \eta_s}
-  2 {\cal T}_{\pi , \eta_s}\ \right]
\nonumber\\
\eta_{\Sigma^+}^{(n), j} & = & 
\left(\ {1\over 2}(5F^2+2DF+D^2)\an + (D-F)^2\bn\ + \rho_{\Sigma^+}^{(n)}
 \right)
\left( L_{ju}-L_\pi\right)
\nonumber\\
& & + {1\over 2} \left( (D-F)^2+{1\over 3}\right)
(\an+4\bn)\left(L_{js}- L_K\right)
\nonumber\\
& & +\ {{\cal C}^2\over 9} \left(\gn-{\sn\over 3}\right)
\left[\ J_\pi + 2 J_K - J_{ju}-2 J_{js}\ \right]
\nonumber\\
\eta_{\Sigma^+}^{(n), r} & = &     
\left(\ {1\over 2}(5F^2+2DF+D^2)\an + (D-F)^2\bn\ 
+ \rho_{\Sigma^+}^{(n)}
\right) 
\left( L_{ru}-L_K\right)
\nonumber\\
& & + {1\over 2}\left( (D-F)^2+{1\over 3}\right) 
(\an+4\bn) \left(L_{rs}-L_{\eta_s}\right)
\nonumber\\
& & +\ {{\cal C}^2\over 9} \left(\gn-{\sn\over 3}\right)
\left[\ J_K + 2 J_{\eta_s} - J_{ru}-2 J_{rs}\ \right]
\nonumber\\
c_{\Sigma^+}^{(n), 0} & = & 
{1\over 6}\overline{m}\left( 4 b_1 + 10 b_2 + b_3 + 2 b_4 + b_5\right)
\ +\ {1\over 6} m_s \left( b_3 - 4 b_4 + 4 b_5\right)
+{1\over 6} (2 b_6+ 5 b_7)\left( 2 m_j+m_r\right)
\nonumber\\
c_{\Sigma^+}^{(n), j} & = & b_8 (m_j-\overline{m})
\nonumber\\
c_{\Sigma^+}^{(n), r} & = & b_8 (m_r-m_s)
\ \ \ .
\label{eq:tSp}
\end{eqnarray}
The functions ${\cal T}_{a , b}$ 
arising from loop graphs involving the {\bf 138} representation 
are shorthand for ${\cal T}_{a , b}={\cal H}(J_a, J_b, J_X)$.
The $\Sigma^-$ matrix elements can be determined from the 
$\Sigma^+$ matrix elements by the replacements
\begin{eqnarray}
w_{\Sigma^-} & = & w_{\Sigma^0}\ =\ w_{\Sigma^+}
\ \ ,\ \ 
\eta_{\Sigma^-}^{(n), j}\ =\ +\eta_{\Sigma^+}^{(n), j}
\ =\ 
\eta_{\Sigma^0}^{(n), j}
\ \ ,\ \ 
\eta_{\Sigma^-}^{(n), r}\ =\ +\eta_{\Sigma^+}^{(n), r}
\ =\ 
\eta_{\Sigma^0}^{(n), r}
\ \ ,
\nonumber\\
c_{\Sigma^-}^{(n), j} &  = & +c_{\Sigma^+}^{(n), j}
\ =\ +c_{\Sigma^0}^{(n), j}
\ \ ,\ \ 
c_{\Sigma^-}^{(n), r}\ =\ +c_{\Sigma^+}^{(n), r}
\ =\ +c_{\Sigma^0}^{(n), r}
\nonumber\\
\rho_{\Sigma^-}^{(n)} & = & -\rho_{\Sigma^+}^{(n)}
\ \ ,\ \ 
\eta_{\Sigma^-}^{(n), 0}\ =\ -\eta_{\Sigma^+}^{(n), 0}
\ \ ,\ \ 
c_{\Sigma^-}^{(n), 0}\ =\ -c_{\Sigma^+}^{(n), 0}
\nonumber\\
\rho_{\Sigma^0}^{(n)} & = & 0
\ \ ,\ \ 
\eta_{\Sigma^0}^{(n), 0}\ =\ 0
\ \ ,\ \ 
c_{\Sigma^0}^{(n), 0}\ =\ 0
\ \ \ .
\label{eq:tSmz}
\end{eqnarray}

The $\Lambda$ the matrix element vanishes when $\overline{\lambda}^3$
is purely isovector, 
however, for arbitrary $y_i$ there is a non-zero contribution,
\begin{eqnarray}
\rho_{\Lambda}^{(n)} & = & 0
\nonumber\\
w_{\Lambda} & = &
\left( -{2\over 3} D^2 + 8 DF - 6 F^2\right) L_\pi
\ +\ 
\left( -{10\over 3} D^2 + 4 DF + 6 F^2\right) L_K
\nonumber\\
& & \ +\ 
\left( {14\over 3} D^2 - 8 DF + 6 F^2\right) (2 L_{ju}+L_{ru})
\ +\ 
{1\over 3}\left(D+3 F\right)^2 (2 L_{js}+L_{rs})
\nonumber\\
& &\ +\ 
{4\over 3} \left( 2D-3F\right)^2 R_{\pi , \pi}
\ -\ 
{4\over 3}\left( 2 D^2 + 3 DF - 9 F^2\right) R_{\pi , \eta_s}
\ +\ 
{1\over 3} \left( D+3F\right)^2 R_{\eta_s , \eta_s}
\nonumber\\
\eta_{\Lambda}^{(n), 0} & = & 0
\nonumber\\
\eta_{\Lambda}^{(n), j} & = & 
\left( {1\over 6}\left( 5D^2-6 DF+9 F^2+ 3\right)\an
+ \left(3\left(D-F\right)^2+1\right)\bn \right)
(L_{ju}-L_\pi)
\nonumber\\ & & 
+\ 
\left(  {1\over 6} \left(D+3F\right)^2 + {1\over 2}\right)\an
(L_{js}-L_K)
\ -\ {{\cal C}^2\over 3} \left(\gn-{\sn\over 3}\right)
\left[\ J_{ju}-J_\pi\ \right]
\nonumber\\
\eta_{\Lambda}^{(n), r} & = &
\left( {1\over 6}\left( 5D^2-6 DF+9 F^2+ 3\right)\an
+ \left(3\left(D-F\right)^2+1\right)\bn \right)
(L_{ru}-L_K)
\nonumber\\ & & 
+\ 
\left(  {1\over 6} \left(D+3F\right)^2 + {1\over 2}\right)\an
(L_{rs}-L_{\eta_s})
\ -\ {{\cal C}^2\over 3} \left(\gn-{\sn\over 3}\right)
\left[\ J_{ru}-J_K\ \right]
\nonumber\\
c_{\Lambda}^{(n), 0} & = & 0
\nonumber\\
c_{\Lambda}^{(n), j} & = & b_8 (m_j-\overline{m})
\nonumber\\
c_{\Lambda}^{(n), r} & = & b_8 (m_r-m_s)
\ \ \  .
\label{eq:tlam}
\end{eqnarray}

For the $\Xi^0$ we find
\begin{eqnarray}
\rho_{\Xi^0}^{(n)} & = & {1\over 6} \left( \an+ 4\bn\ \right)
\nonumber\\
w_{\Xi^0} & = & 
-2(D^2-6DF+3F^2) L_K + 2 (3F^2-D^2)L_{\eta_s} 
\nonumber\\
& & +3(D-F)^2 (2L_{ju}+L_{ru}) + 2 (D^2+3F^2)(2 L_{js}+L_{rs})
\nonumber\\
& & + 3 \left[\ 4 F^2 R_{\eta_s , \eta_s} + 4 F(F-D) R_{\pi , \eta_s}
+ (D-F)^2 R_{\pi , \pi}\ \right]
\nonumber\\
& & +\ {{\cal C}^2\over 3} \left[\ 5 J_K + J_{\eta_s}+4 J_{ju}+2J_{ru}
+2 J_{js} + J_{rs}\ \right]
 +\ {2{\cal C}^2\over 3}  
\left[\ {\cal T}_{\pi , \pi} +  {\cal T}_{\eta_s , \eta_s}
-  2 {\cal T}_{\pi , \eta_s}\ \right]
\nonumber\\
\eta_{\Xi^0}^{(n), 0} & = & 
\left[\ (D^2+4 DF - F^2)\an + 4 F (D-F)\bn\ \right] L_K
\nonumber\\
& & + \left[\ (F^2-D^2) \an + 4 F^2 \bn\  \right] L_{\eta_s}
\nonumber\\
& & + \left[\  (D^2+F^2)\an + 4F^2\bn\ \right] 
\left(\ 2 L_{js} + L_{rs}\ \right)
\nonumber\\
& & -\  \rho_{\Xi^0}^{(n)} (2 L_{ju}+L_{ru})
\nonumber\\
& & + 3 \rho_{\Xi^0}^{(n)}\left[\ 
4 F^2 R_{\eta_s , \eta_s} + 4 F(F-D) R_{\pi , \eta_s}
+ (D-F)^2 R_{\pi , \pi}\ \right]
\nonumber\\
& & -\ {{\cal C}^2\over 9} \left(\gn-{\sn\over 3}\right)
\left[\ 2 J_K+J_{\eta_s} + 2 J_{js}+J_{rs}\ \right]
\nonumber\\
& & -\ {2{\cal C}^2\over 9} \left(\gn-{\sn\over 3}\right)
\left[\ {\cal T}_{\pi , \pi} +  {\cal T}_{\eta_s , \eta_s}
-  2 {\cal T}_{\pi , \eta_s}\ \right]
\nonumber\\
\eta_{\Xi^0}^{(n), j} & = & 
3 (D-F)^2  \rho_{\Xi^0}^{(n)} \left( L_{ju}-L_\pi\right)
\nonumber\\
& & 
+\ \left[\ {1\over 2}(D^2+2 DF+5F^2)\an + (D-F)^2\bn\ \right] 
\left(L_{js}-L_K\right)
\nonumber\\
& & + \ \rho_{\Xi^0}^{(n)} (L_{ju}-L_\pi)
+{1\over 6}(5\an+2\bn) (L_{js}-L_K)
\nonumber\\
& & +\ {{\cal C}^2\over 9} \left(\gn-{\sn\over 3}\right)
\left[\ 2 J_\pi+J_K-2 J_{ju}-J_{js}\ \right]
\nonumber\\
\eta_{\Xi^0}^{(n), r} & = & 
3 (D-F)^2 \rho_{\Xi^0}^{(n)} \left( L_{ru}-L_K\right)
\nonumber\\
& & 
+\ \left[\ {1\over 2}(D^2+2 DF+5F^2)\an + (D-F)^2\bn\ \right] 
\left(L_{rs}-L_{\eta_s}\right)
\nonumber\\
& & + \ \rho_{\Xi^0}^{(n)} (L_{ru}-L_K)
+{1\over 6}(5\an+2\bn) (L_{rs}-L_{\eta_s})
\nonumber\\
& & +\ {{\cal C}^2\over 9} \left(\gn-{\sn\over 3}\right)
\left[\ 2 J_K+J_{\eta_s}-J_{rs}-2 J_{ru}\ \right]
\nonumber\\
c_{\Xi^0}^{(n), 0} & = & {1\over 3}\overline{m}\left( 4 b_1+b_2\right)
\ +\ {1\over 6} m_s \left( 4 b_3-4 b_4 + b_5\right) 
+ {1\over 6} ( 4 b_6+b_7)  \left(2 m_j+m_r\right)
\nonumber\\
c_{\Xi^0}^{(n), j} & = & b_8 (m_j-\overline{m})
\nonumber\\
c_{\Xi^0}^{(n), r} & = & b_8 (m_r-m_s)
\ \ \  .
\label{eq:tCz}
\end{eqnarray}
The matrix elements for the $\Xi^-$ can be found by the replacements
\begin{eqnarray}
w_{\Xi^-} & = & w_{\Xi^0}
\ \ ,\ \ 
\eta_{\Xi^-}^{(n), j}\ =\ +\eta_{\Xi^0}^{(n), j}
\ \ ,\ \ 
\eta_{\Xi^-}^{(n), r}\ =\ +\eta_{\Xi^0}^{(n), r}
\ \ ,\ \ 
c_{\Xi^-}^{(n), j}\ =\ +c_{\Xi^0}^{(n), j}
\ \ ,\ \ 
c_{\Xi^-}^{(n), r}\ =\ +c_{\Xi^0}^{(n), r}
\nonumber\\
\rho_{\Xi^-}^{(n)} & = & -\rho_{\Xi^0}^{(n)}
\ \ ,\ \ 
\eta_{\Xi^-}^{(n), 0}\ =\ -\eta_{\Xi^0}^{(n), 0}
\ \ ,\ \ 
c_{\Xi^-}^{(n), 0}\ =\ -c_{\Xi^0}^{(n), 0}
\ \ \ .
\label{eq:tCm}
\end{eqnarray}

It is straightforward to show that the divergences occurring in
eqs.~(\ref{eq:tp}), (\ref{eq:tSp}), (\ref{eq:tlam})
and (\ref{eq:tCz}) (and hence the 
entire multiplet) can be absorbed by the 
eight counterterms $b_1,... b_8$.
Further, in the QCD limit, where $m_j\rightarrow\overline{m}$ and
$m_r\rightarrow m_s$, the matrix elements become 
independent of  the choice of charges in the sea- and ghost-quark sectors,
and one recovers the matrix elements of QCD with a heavy $\eta^\prime$.

As is the case for the magnetic moments, one can determine values of the 
charges $y_{jl}$ and $y_r$ that minimize
the one-loop contribution from the sea-quarks.
It is conceivable that this choice of charges will  minimize 
the uncertainty in the determination of the 
leading order matrix elements in eq.~(\ref{eq:tree}), and the counterterms
in eq.~(\ref{eq:tcts}).
However,  it may be the case that consideration of several different
charge combinations will provide the most useful information.

\section{Conclusions}
We have included the lowest-lying octet- and decuplet-baryons into 
partially quenched chiral perturbation theory.  
In addition to the octet- and decuplet- baryons formed from the three 
light valence quarks, baryons containing ghost-quarks and 
baryons containing sea-quarks (and those containing both) have been
included to accomplish the partial quenching.
The leading one-loop contributions to the octet-baryon masses, 
magnetic moments
and matrix elements of isovector twist-2 operators are presented.

The extension of both the electric charge matrix and the isovector charge
matrix into the sea- and ghost-sectors is not unique, constrained only by the 
requirement that QCD is recovered
in the limit where the sea-quark masses are equal to the
valence quark masses.
This leads to additional freedom and couplings 
in PQQCD for the magnetic moments and 
matrix elements of isovector twist-2 operators.
These extra couplings are directly related to the contribution 
of disconnected quark-line diagrams to these observables.
It is hoped that this extra freedom can be exploited to study the 
chiral expansion and to reduce the uncertainty in
chiral extrapolations.

\bigskip\bigskip

\acknowledgements

We would like to thank Steve Sharpe for many useful discussions and to
acknowledge his involvement in the early stages of this work.
We also thank Maarten Golterman for useful discussions.
MJS is supported in
part by the U.S. Dept. of Energy under Grant No.  DE-FG03-97ER4014
and JWC is supported in
part by the U.S. Dept. of Energy under Grant No. DE-FG02-93ER-40762.

\end{document}